\documentclass[nonacm]{acmart}

\usepackage[linesnumbered,ruled,vlined,longend,norelsize]{algorithm2e}
\usepackage{todonotes}
\usepackage{subfig}
\usepackage{siunitx}
\usepackage{array}
\usepackage{multirow}
\usepackage{multicol}
\usepackage{pdflscape}
\usepackage{diagbox}
\usepackage{placeins} 

\graphicspath{{img/}}

\renewcommand\footnotetextcopyrightpermission[1]{} 

\title{Structural Tree Extraction from 3D Surfaces}

\author{Diogo de Andrade}
\email{diogo.andrade@ulusofona.pt}
\orcid{0000-0001-9180-0545}
\affiliation{%
  \institution{Lusófona University, HEI‐Lab: Digital Human‐Environment Interaction Labs}
  \streetaddress{Campo Grande, 376}
  \city{Lisboa}
  \country{Portugal}
  \postcode{1749-024}
}

\author{Nuno Fachada}
\email{nuno.fachada@ulusofona.pt}
\orcid{0000-0002-8487-5837}
\affiliation{%
  \institution{Lusófona University, COPELABS}
  \streetaddress{Campo Grande, 376}
  \city{Lisboa}
  \country{Portugal}
  \postcode{1749-024}
}
\affiliation{%
  \institution{Center of Technology and Systems (UNINOVA-CTS) and Associated Lab of Intelligent Systems (LASI)}
  \city{Caparica}
  \country{Portugal}
  \postcode{2829-516}
}

\begin{document}

\begin{abstract}
This paper introduces a method to extract a hierarchical tree representation from 3D unorganized polygonal data. The proposed approach first extracts a graph representation of the surface, which serves as the foundation for structural analysis. A Steiner tree is then generated to establish an optimized connection between key terminal points, defined according to application-specific criteria. The structure can be further refined by leveraging line-of-sight constraints, reducing redundancy while preserving essential connectivity. Unlike traditional skeletonization techniques, which often assume volumetric interpretations, this method operates directly on the surface, ensuring that the resulting representation remains relevant for navigation-aware geometric analysis.
The method is validated through two use cases: extracting structural representations from tile-based elements for procedural content generation, and identifying key points and structural metrics for automated level analysis. Results demonstrate its ability to produce simplified, coherent representations, supporting applications in procedural generation, spatial reasoning, and map analysis.
\end{abstract}

\maketitle

\section{Introduction}

Triangle meshes are widely used to represent the geometry of 3D environments. However, their complexity--especially in non-trivial scenarios--introduces significant challenges for computational applications such as navigation, level analysis, sound propagation, and spatial optimization~\cite{snook2000navmesh,lamarche2004crowd, tagliasacchi20163d}. These tasks often require simplified representations that preserve essential structural properties while reducing computational overhead. Conventional methods--such as medial axis transformations, skeletonization, and sampling-based approaches--have been employed to derive such simplified outlines. However, these techniques exhibit various limitations, including instability in 3D settings~\cite{dey2006defining}, sensitivity to noise~\cite{tagliasacchi20163d}, and the frequent necessity of parameter tweaking, involving substantial expertise~\cite{lazarus1999level}.

This work introduces a method for extracting a hierarchical structure from polygonal data, enabling efficient representation of surfaces while retaining critical features for navigation and analysis. Central to the proposed approach is the use of a Steiner tree~\cite{kou1981fast}, which provides a minimal yet spatially coherent layout by optimizing connections between key regions of the surface, preserving structural highlights while reducing computational overhead. The Steiner tree is constructed over a surface-derived graph and then simplified based on line-of-sight (LoS) constraints.  In contrast to many skeletonization techniques that assume a volumetric interpretation of geometry, the proposed approach operates directly on the surface, which is advantageous for various practical applications such as procedural content generation, spatial analysis in virtual environments, layout evaluation in game levels,  and navigation-aware geometry deformation.

\begin{figure}[tb]
    \centering
    \includegraphics[width=.6\textwidth]{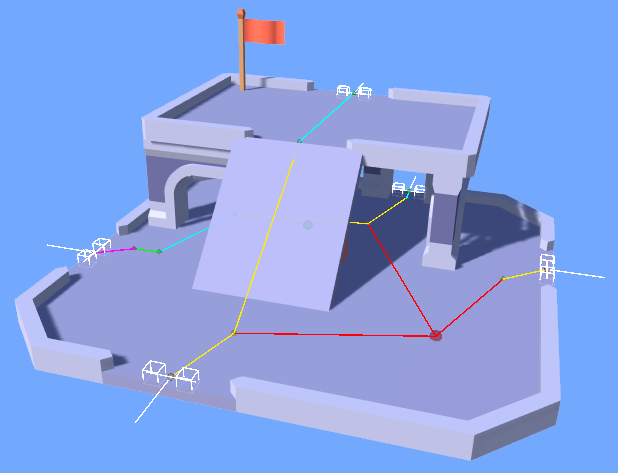}
    \Description{Structure extraction from a polygon mesh. The extracted graph represents navigable paths between the various entrances/exits of the mesh. The red sphere represents the root of the tree structure, while the colors represent the depth of the tree (red is depth 1, yellow is depth 2, and so forth).}
    \caption{Structure extraction from a polygon mesh. The extracted graph represents navigable paths between the various entrances/exits of the mesh. The red sphere represents the root of the tree structure, while the colors represent the depth of the tree (red is depth 1, yellow is depth 2, and so forth).}
    \label{fig:structure}
\end{figure}

\begin{figure}[tb]
    \centering
    \includegraphics[width=\textwidth]{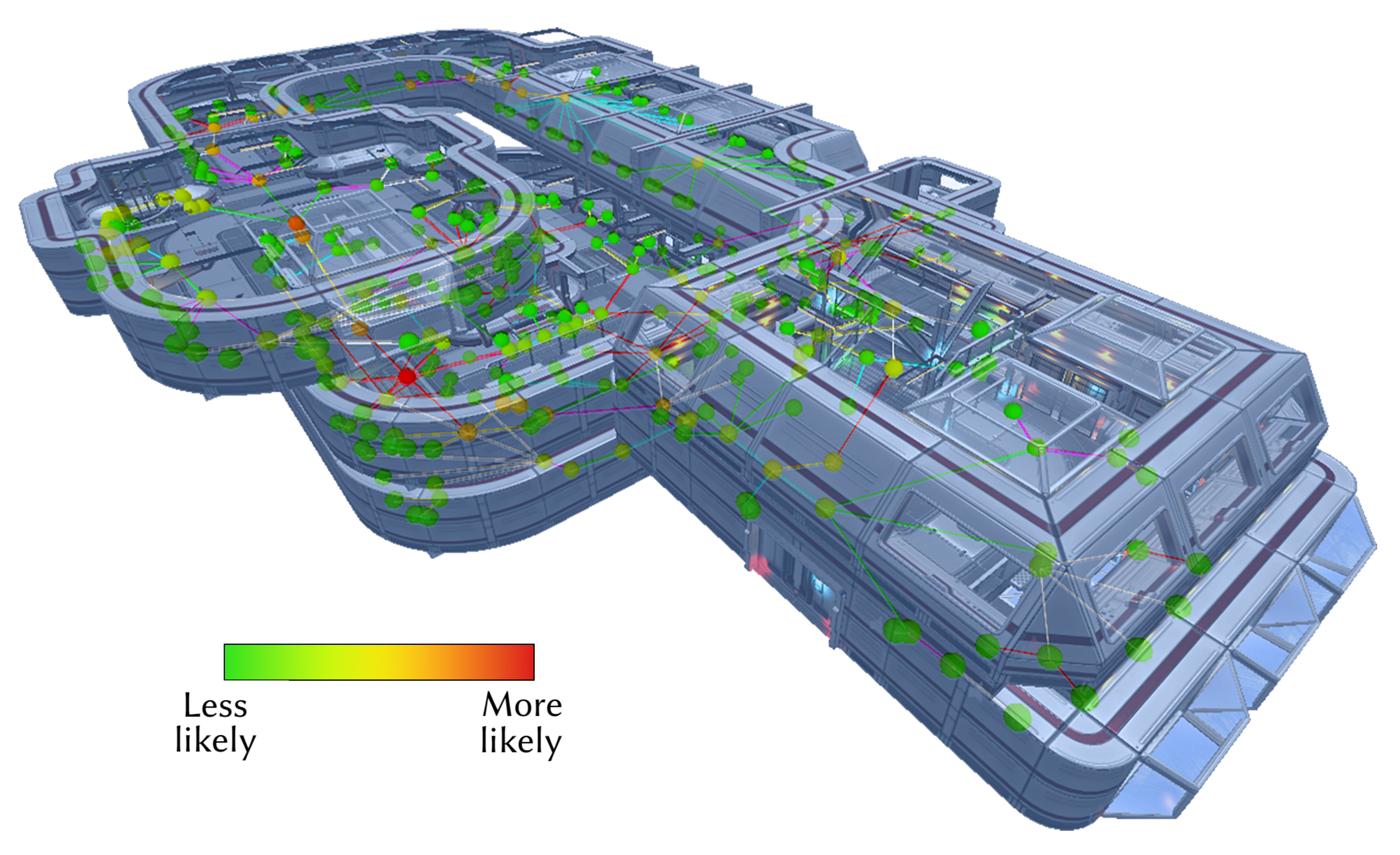}
    \caption{Map analysis: node color gradient indicates the likelihood of player traversal across different areas of the level.}
    \Description{A large 3D map of a spaceship that could be part of a videogame, with a tree-like structure with colored nodes that show the probability of passing through that section.}
    \label{fig:map_analysis}
\end{figure}

For evaluation purposes, the method is applied in two scenarios. The first consists of deriving navigable structures from individual geometry elements, as illustrated in Fig.~\ref{fig:structure}. The second, depicted in Fig.~\ref{fig:map_analysis}, is oriented towards extracting metrics and identifying points of interest in a game level. In both scenarios, the method yields coherent and simplified representations of the input geometry. In the tile-based case, it preserves primary navigational routes while reducing structural complexity. In the level analysis, it highlights areas of potential traversal density through centrality measures, offering a compact basis for spatial reasoning, and automated evaluation.

The remainder of this article is organized as follows: Section~\ref{sec:related-work} reviews related work on structural extraction techniques, Steiner trees, and automated level analysis. The proposed method, its implementation, and evaluation scenarios are detailed in Section~\ref{sec:method}. Section~\ref{sec:results} presents the experimental results, as well as the evaluation of the proposed approach in different scenarios. Section~\ref{sec:discussion} discusses limitations and future improvements, while Section~\ref{sec:conclusion} provides concluding remarks.

\section{Related Work}
\label{sec:related-work}

The extraction of structural representations from polygonal meshes has been widely studied across several domains~\cite{tagliasacchi20163d,au2008skeleton,lamarche2004crowd,bespalov2003reeb}. The approach proposed in this paper intersects with three areas of research: 1) skeleton extraction techniques, which aim to derive compact and meaningful structures from geometric data; 2) Steiner tree formulations, which offer optimized representations of connectivity in graph-based models; and, 3) automated level analysis, which leverages structural representations to evaluate spatial properties of game environments, supporting tasks such as navigation, design evaluation, and procedural content generation (PCG).

\subsection{Skeleton extraction}

The extraction of structural representations from polygonal or triangular meshes has several applications in computational geometry, including shape analysis, animation, or navigation~\cite{tagliasacchi20163d}. Various approaches have been developed to derive skeletal structures, often with different objectives and underlying assumptions.

The medial axis transformation, introduced by Lee~\cite{lee1982medial}, provides a curved skeleton representation of planar shapes. Aichholzer et al.~\cite{aichholzer1995novel} proposed the straight skeleton alternative, which provides a representation more amenable to graph-based and tree-based structures. However, these methods suffer from instability, particularly in 3D environments~\cite{dey2006defining}.

Mesh contraction techniques, such as those discussed by Au et al.~\cite{au2008skeleton}, generate skeletal structures by iteratively collapsing edges. These methods have found extensive use in skeletal deformation modeling and character animation.
Reeb graphs, on the other hand, encode topological changes in scalar functions defined over a surface~\cite{reeb1946points}, and have applications across various domains~\cite{shinagawa1991surface,bespalov2003reeb,shailja2021computational, pascucci2007robust}. Extensions to Reeb graphs, such as level set diagrams~\cite{lazarus1999level}, further refine structural extraction by explicitly identifying structural loops within a surface.
Another distinct class of methods employs potential fields, in which scalar functions encode notions of connectivity and structural salience. For instance, Chuang et al.~\cite{chuang2000skeletonisation} introduced a skeletonization method based on electrostatic potential fields, effectively capturing structural characteristics for geometric analysis.

While effective for geometric abstraction, skeletonization methods typically operate directly on the mesh and often assume a volumetric interpretation. Even surface-based variants, such as level set diagrams, rely on medial structures or loop detection that infer an internal structure from surface properties, making them ill-suited for cases where the primary interest lies in surface-level structural detail, such as navigation analysis.

\subsection{Steiner trees}

While skeletonization techniques typically derive structural representations directly from a volumetric or surface interpretation, Steiner tree formulations emphasize connectivity between discrete elements. Steiner trees operate over abstract graph structures, generating minimal spanning frameworks that represent essential relational properties rather than geometric form.

The Steiner tree problem, defined over a graph, seeks a minimum-cost tree that connects a specified subset of nodes—referred to as \emph{terminals}—while optionally including additional intermediate nodes, known as \emph{Steiner points}, selected from the existing nodes in the graph. Due to its computational complexity, shown to be NP-complete~\cite{book1975richard}, exact solutions are impractical for large graphs, leading to the development of heuristic and approximation algorithms~\cite{kou1981fast, hougardy19991, robins2000improved, byrka2010improved}. These methods have been widely applied in network design, circuit layout optimization, and computational geometry~\cite{ljubic2021solving}.
The present work employs Steiner tree approximations to construct an efficient and minimal representation of a geometric surface, incorporating domain-specific constraints to ensure relevance to navigation and related tasks.

\subsection{Automated level analysis}

Automated analysis of game levels emerged as a response to the limitations of manual inspection and playtesting in evaluating spatial design. Early work on navigation meshes and spatial reasoning for virtual environments emphasized the encoding of traversability and connectivity in structured representations~\cite{snook2000navmesh, lamarche2004crowd}. These approaches laid the groundwork for understanding spatial flow and accessibility. With the rise of procedural content generation (PCG), the demand for tools capable of evaluating large volumes of automatically generated content increased substantially~\cite{shaker2016procedural}. This led to more formalized approaches to structural evaluation, including methods based on graph metrics~\cite{liapis2013towards}, spatial decomposition~\cite{lamarche2004crowd}, and simulation-based analysis~\cite{shaker2010towards}.

Later work explored the integration of automated analysis into mixed-initiative content creation (MICC), where it informs and reacts to designer input~\cite{liapis2013towards}. In this context, structural features extracted from levels are used to validate gameplay properties, generate design feedback, and steer algorithmic decisions. 
The contribution of the present work lies in the extraction of a simplified structure from 3D navigable surfaces using a Steiner tree approximation built over a surface-derived graph. This representation captures essential spatial relationships while reducing topological complexity. Once extracted, it can be used to support downstream tasks such as layout evaluation, efficient computation of structural metrics, and identification of gameplay-relevant landmarks.

\section{Methodology}
\label{sec:method}

This section presents the algorithm for extracting a simplified representation from unstructured geometry. The method constructs one or more hierarchical trees over a graph derived from the input surface, enabling spatial abstraction suitable for downstream use and/or analysis. Subsection~\ref{sec:pipeline} outlines the main stages of the algorithm, while Subsection~\ref{sec:implementation} discusses technical implementation details. Subsection~\ref{sec:validation} describes the procedures used to validate the approach.

\subsection{Tree structure extraction pipeline}
\label{sec:pipeline}

\begin{figure}[tb]
    \centering
    \Description{The different steps on structure extraction.}
    \subfloat[]{
        \includegraphics[width=0.2\textwidth]{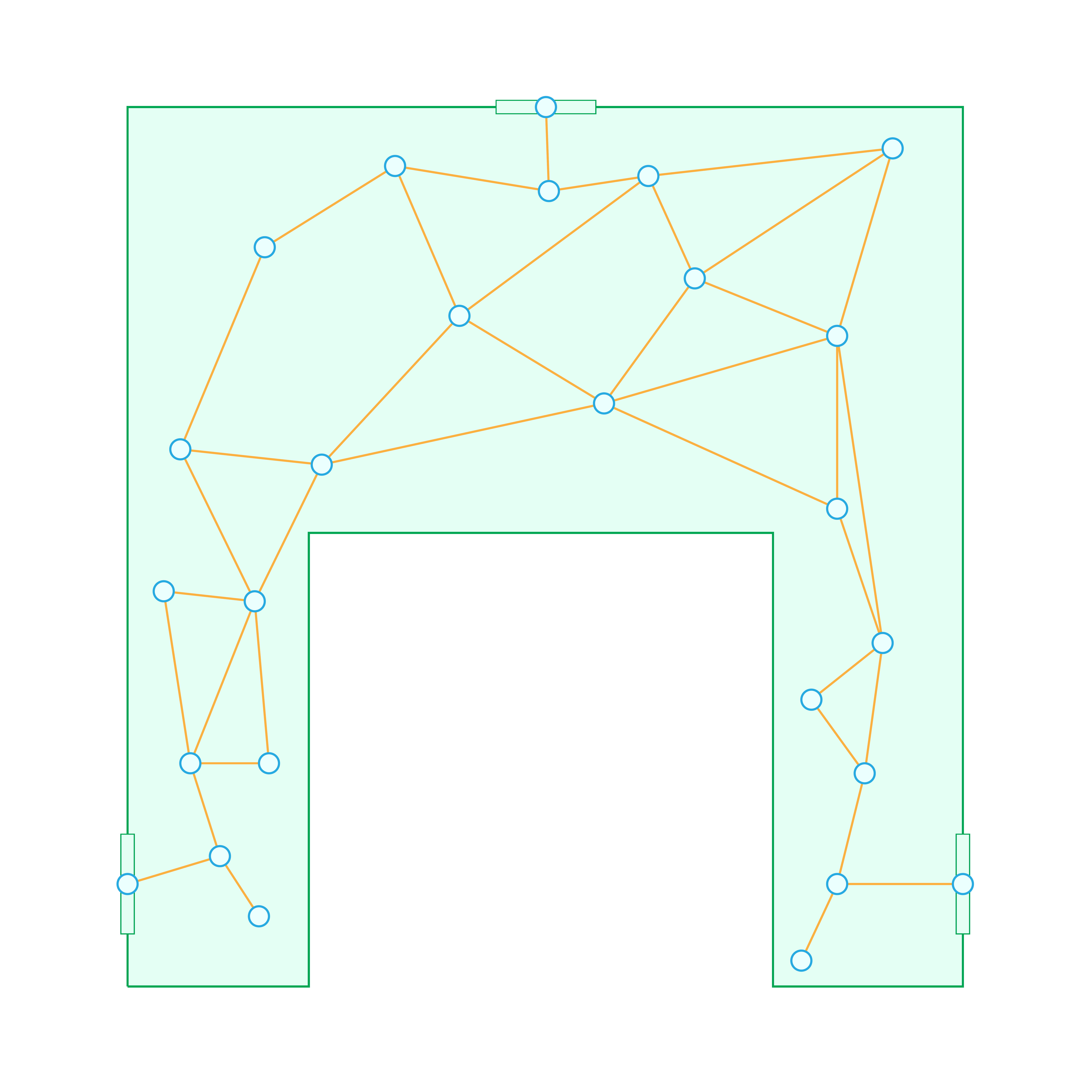}
        \label{fig:sdiag01}}
    \subfloat[]{
        \includegraphics[width=0.2\textwidth]{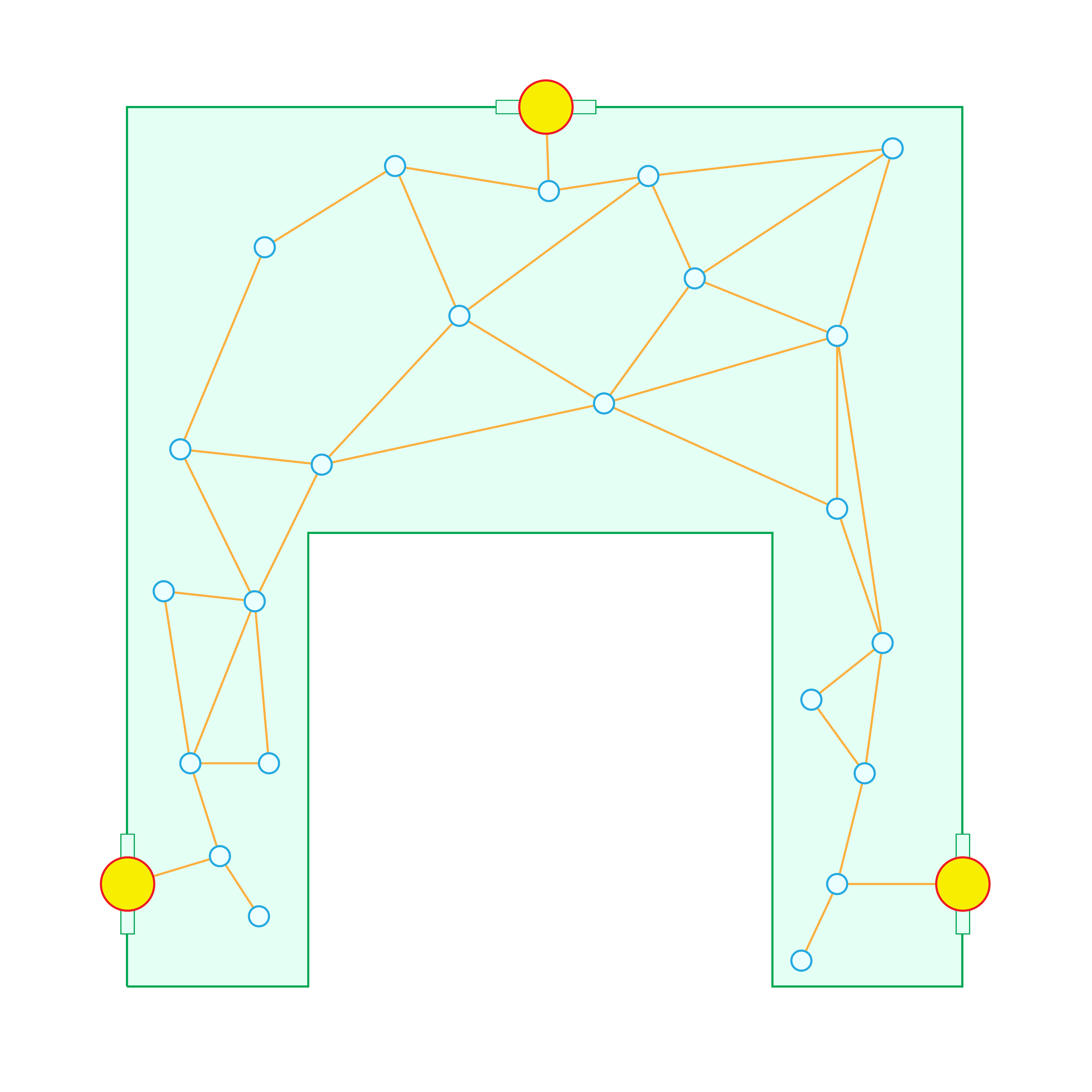}
        \label{fig:sdiag02}}
    \subfloat[]{
        \includegraphics[width=0.2\textwidth]{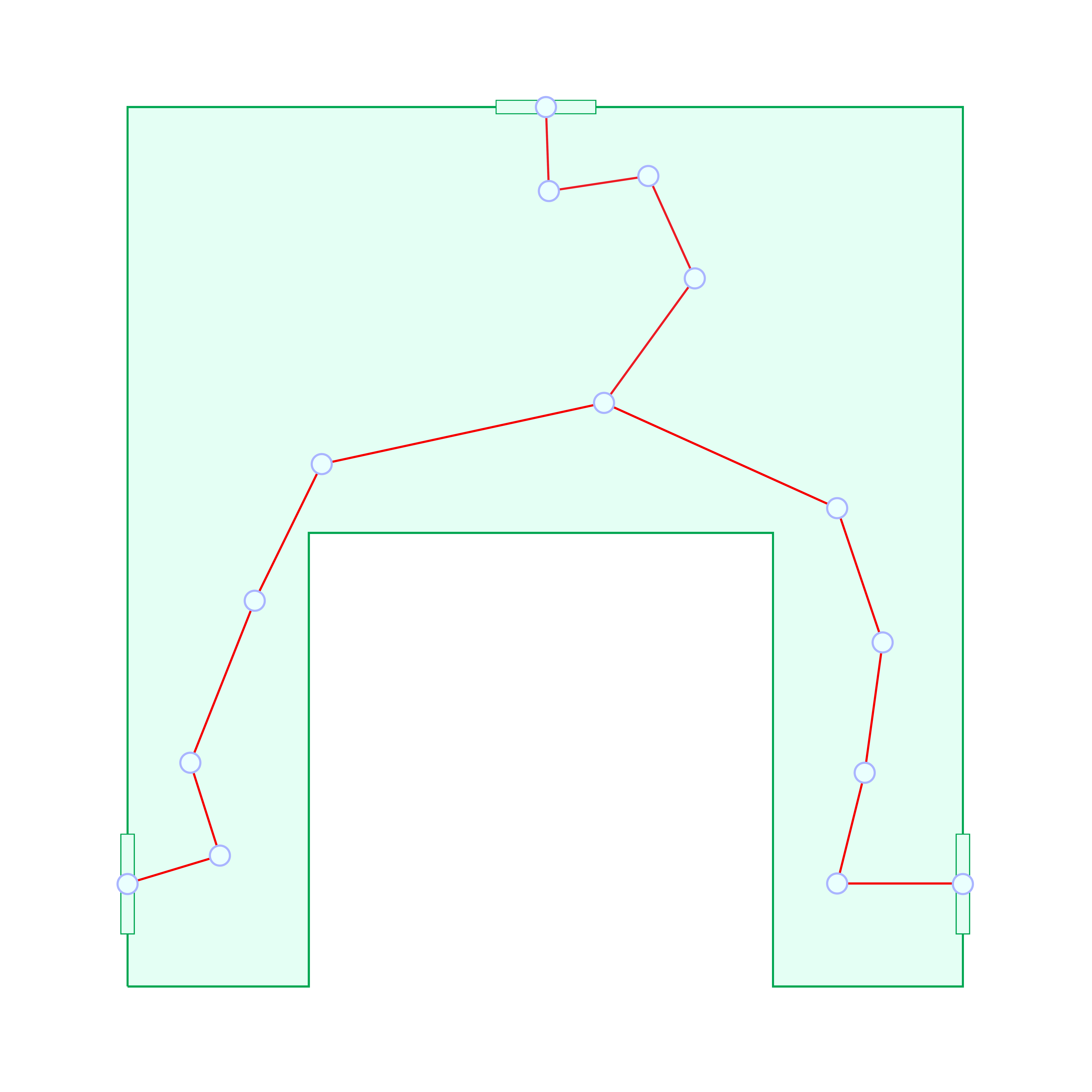}
        \label{fig:sdiag03}}
    \subfloat[]{
        \includegraphics[width=0.2\textwidth]{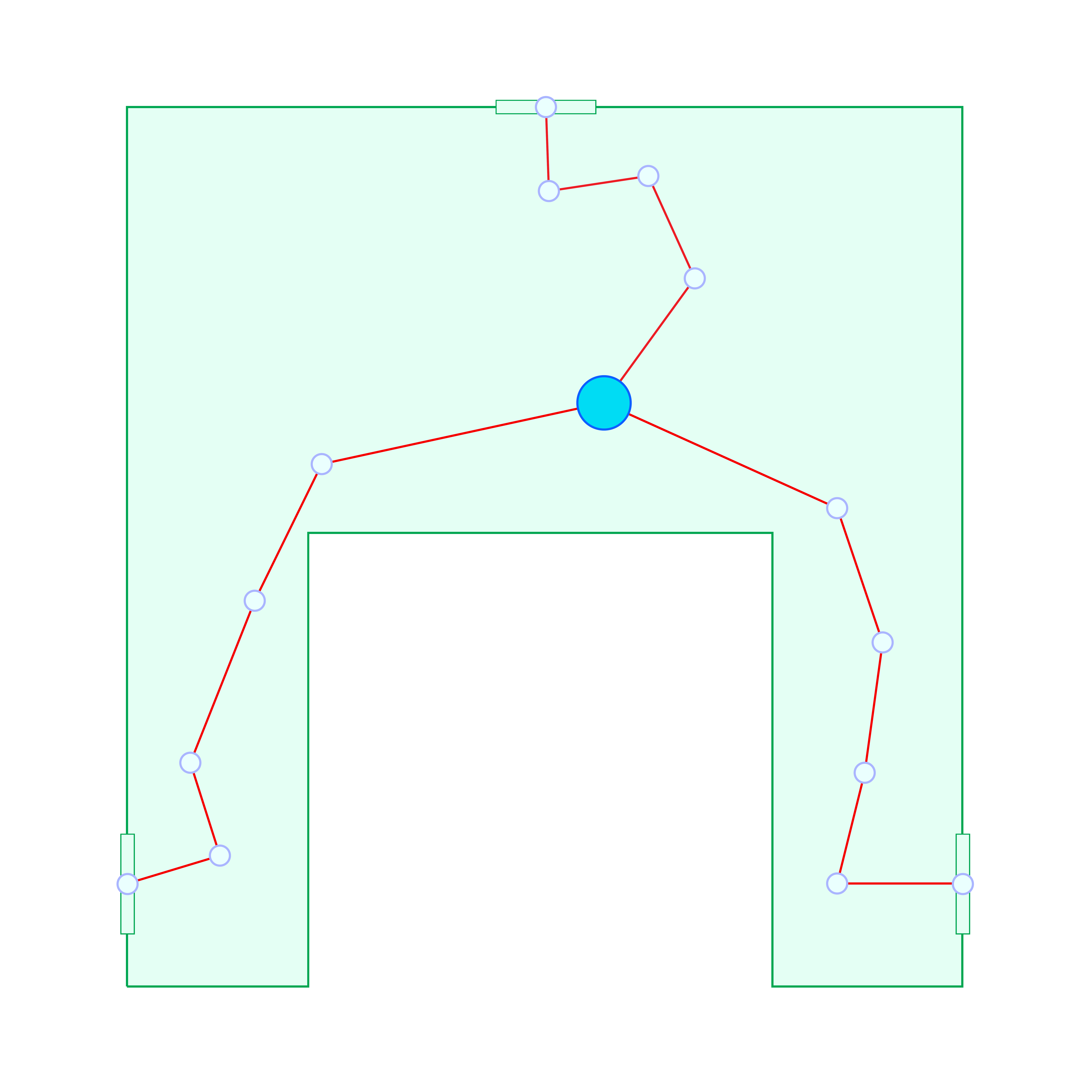}
        \label{fig:sdiag04}}
    \subfloat[]{
        \includegraphics[width=0.2\textwidth]{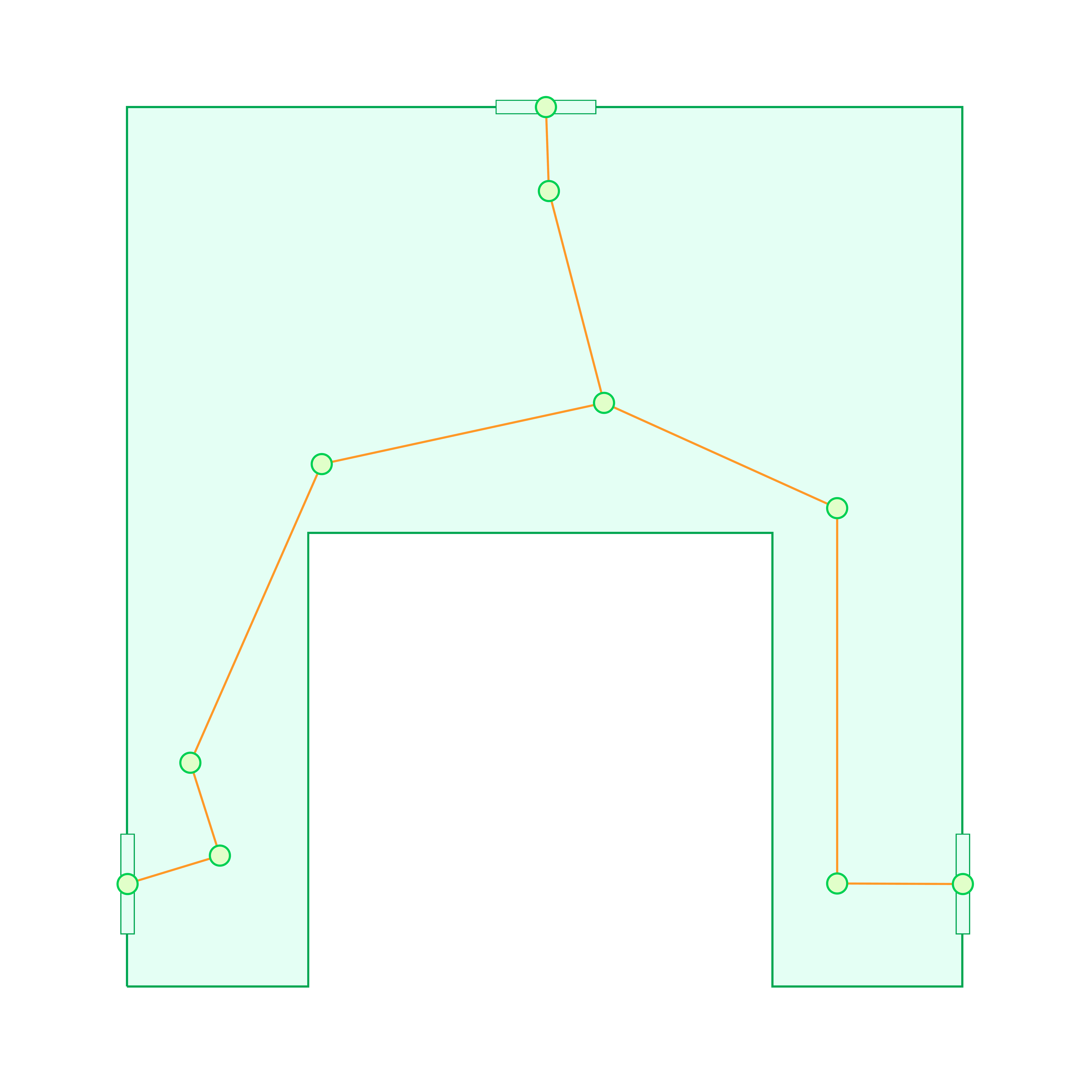}
        \label{fig:sdiag05}}
    \caption{Algorithm~\ref{alg:gen_structure} steps: a) build initial graph; b) identify terminal nodes; c) build Steiner tree; d) find rooted trees; and, e) simplify trees.}
    \label{fig:alg_steps}
\end{figure}

The proposed method can be visualized in Fig.~\ref{fig:alg_steps} and formalized as Algorithm~\ref{alg:gen_structure}. First, a graph representation is extracted from the input data, defining a search space for path connectivity. In the following step, terminal points, which serve as structurally relevant locations, are identified based on application-specific criteria. A Steiner tree is then constructed, forming an optimized connection between these points. In the next step, the unrooted Steiner tree is converted into one or more rooted trees using a centrality-based criterion to establish one or more root nodes. Finally, the resulting directed trees are simplified using LoS constraints.

\begin{algorithm}[tb]
  \DontPrintSemicolon
  \caption{Tree structure extraction from unstructured triangular meshes.\label{alg:gen_structure}.}

  \SetKwData{Geometry}{geometry}
  \SetKwData{Graph}{graph}
  \SetKwData{Trees}{trees}
  \SetKwData{SteinerTree}{steinerTree}
  \SetKwData{TerminalNodes}{terminalNodes}
  \SetKwData{SimplifiedTrees}{simplifiedTrees}
  \SetKwData{Trees}{trees}

  \SetKwFunction{BuildInitialGraph}{BuildInitialGraph}
  \SetKwFunction{IdentifyTerminalNodes}{IdentifyTerminalNodes}
  \SetKwFunction{FindRoot}{FindRoot}
  \SetKwFunction{BuildSteinerTree}{BuildSteinerTree}
  \SetKwFunction{SimplifyTree}{SimplifyTree}
  \SetKwFunction{FindRootedTrees}{FindRootedTrees}

  \SetKwInOut{Input}{Input}
  \SetKwInOut{Output}{Output}

  \Input{\Geometry}
  \Output{\SimplifiedTrees}

  \BlankLine

  \Graph $\leftarrow$ \BuildInitialGraph{\Geometry} \label{alg:init_graph} \\
  \TerminalNodes $\leftarrow$ \IdentifyTerminalNodes{\Graph} \label{alg:terminal_nodes} \\
  \SteinerTree $\leftarrow$ \BuildSteinerTree{\Graph, \TerminalNodes} \label{alg:steiner_tree} \\
  \Trees $\leftarrow$ \FindRootedTrees{\SteinerTree} \label{alg:find_roots} \\
  \SimplifiedTrees $\leftarrow$ \SimplifyTree{\Trees} \label{alg:simplify_tree}

\end{algorithm}

\subsubsection{Building the initial graph}
\label{sec:initial_graph}

The initial graph is a superset of all potential solution graphs, defining the search space. The method of constructing this graph depends on the input data and the specific use case. The following subsections describe two such methods: one based on extracting connectivity from a navigation mesh and another utilizing a sampling-based strategy.

\paragraph{Initial graph from navigation mesh}

When the objective of the process is related to agent movement and/or map analysis, the geometry's navigation mesh can directly provide the required graph-like structure. More specifically, each node in the initial graph corresponds to a single convex planar polygon within the navigation mesh. The edges (or links) of the graph are undirected and extracted from the adjacency relationships among these polygons. The weight of the links is determined by the Euclidean distance between the associated node positions.

Although this construction aligns naturally with navigation-related objectives, it also imposes certain limitations. Since the navigation mesh only encodes walkable areas, the resulting graph exclusively captures structure in terms of agent walkability. As a consequence, any features of the geometry that are not navigable--such as overhangs, verticality, or fine structural elements disconnected from the ground--are omitted from the representation. Furthermore, because each polygon is treated as a single node regardless of its size, large flat areas are represented in the same way as small detailed regions. This uniform treatment may lead to distortions in structural representativity, particularly in contexts where geometric complexity or spatial density is of greater interest than navigability.

\paragraph{Initial graph from sampling}

In cases where navigation structures are not available or a higher level of granularity is required, an initial graph can be constructed through a sampling process applied to the source data. Each sampled point constitutes a node, with connectivity determined by a predefined neighborhood structure, such as regular or irregular grids, or a data-driven adjacency criterion. Edge weights can be assigned based on relevant properties of the sampled data--such as geometric features, material characteristics, or visibility relationships--using metrics like Euclidean distance or LoS constraints. This allows the resulting graph to capture structural variations that support further analysis or optimization.

\subsubsection{Identifying terminal nodes}
\label{sec:terminal_nodes}

The second step of Algorithm~\ref{alg:simplify_tree} consists of identifying terminal nodes in the initial graph. These nodes constitute key points within the initial graph that must be retained in the final solution. The selection of terminal nodes depends on the application context, with different scenarios having specific prerequisites. Two methods for determining terminal nodes are described below: identifying entry/exit points of the geometry and leveraging graph-theoretic metrics.

\paragraph{Entry/exit points}

\begin{figure}[tb]
    \centering
    \includegraphics[width=.5\textwidth]{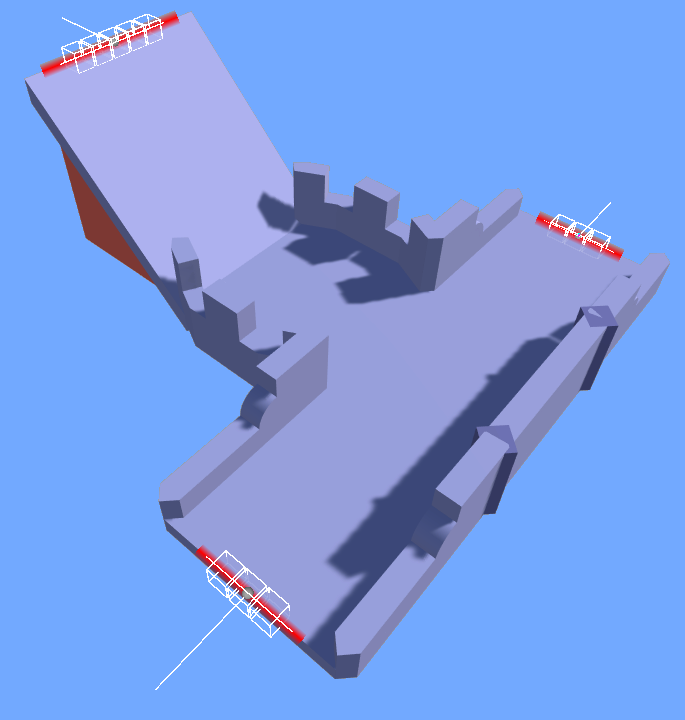}
    \caption{Map piece with entry/exit point.}
    \Description{A geometry map piece with entry/exit points showing in red}
    \label{fig:entry_exit}
\end{figure}

The identification of entry/exit points within a given geometry, depicted in Fig.~\ref{fig:entry_exit}, is performed with the process described by de Andrade and Fachada~\cite{de2023automated}, primarily designed for procedural level generation through geometry elements called ``map pieces''.

The method requires a navigation mesh to identify the entry/exit points. This navigation mesh can be built just for this purpose, or an existing one can be reused. The algorithm starts by extracting the boundary edges of that navigation mesh, which are then sampled at regular intervals. Each sampled boundary point is classified as a potential segment start, potential segment end, both, or neither.
This classification is based on a local raycasting procedure: for each point, a series of raycasts are performed within a fixed radius, oriented along both the forward and backward directions of the boundary (defined as clockwise or counterclockwise with respect to the boundary traversal). If none of the raycasts in the forward direction produce a hit, the point is marked as a potential segment start; if none of the raycasts in the backward direction produce a hit, it is marked as a potential segment end. Points are classified as both if raycasts in neither direction result in a hit.

After classification, the algorithm iterates along the boundary to extract segments. When a potential start point is encountered, a segment is initialized and extended forward along the boundary. The extension continues until one of two conditions is met: 1) an end point is followed by a point that is not marked as a potential end, or, 2) the turning angle between consecutive points exceeds a specified sharpness threshold. For each extracted boundary segment, the corresponding segment in the original geometry is then identified using a heuristic matching function that evaluates geometric similarity and spatial proximity. Once a match is established, the segment endpoints are registered as potential entry or exit points and are added to the graph, connected to the nearest existing nodes. These new nodes are subsequently used as terminal nodes in the Steiner tree generation process.

\paragraph{Graph metrics}
\label{sec:graph_metrics}

Graph-theoretic metrics can also be used to identify terminal nodes within the graph. Several of these metrics are summarized in Table~\ref{tab:graph_metrics}.

Degree centrality~\cite{kenett2015networks} quantifies the number of edges connected to a node, making it useful for identifying hubs with varying levels of connectivity. Katz centrality~\cite{katz1953new} extends this concept by incorporating indirect connections, where an attenuation factor controls the influence of distant nodes, thereby weighting closer connections more heavily. Betweenness centrality~\cite{freeman1977set} measures how frequently a node lies on the shortest paths between other nodes, identifying structural junctions that function as bridges. Eigenvector centrality~\cite{bonacich1972factoring} assigns higher scores to nodes connected to other highly ranked nodes, capturing globally influential regions within the graph. A wide range of additional metrics~\cite{saxena2020centrality} may also be applied, depending on the structural emphasis required.

For example, in scenarios that emphasize boundary or endpoint features, selecting leaf nodes (i.e., nodes of degree one) may be sufficient. In contrast, when the structure is interpreted in terms of potential flow, congestion, or interaction--such as in navigation, surveillance, or conflict-prone regions—--centrality-based metrics such as betweenness or Katz centrality provide more informative guidance. High betweenness nodes may indicate choke points or frequently traversed corridors, while high Katz scores can reveal regions of broader influence, both of which contribute to the selection of semantically meaningful terminals.

The choice of metric should reflect the structural assumptions of the domain and the behavioral or functional properties of interest. In practice, hybrid strategies--such as filtering by degree and ranking by centrality--can help mitigate the biases inherent in any single metric, especially in irregular or noisy graphs.

\begin{table}[h]
    \centering
    \caption{Graph metrics used for selecting terminal nodes. For \textbf{degree centrality} $C_D(v)$, $v$ is a node, $\deg(v)$ is the number of direct connections to that node. In \textbf{Betweenness Centrality}, $\sigma_{st}$ is the number of paths between $s$ and $t$ and $\sigma_{st}(v)$ is the number of paths between $s$ and $t$ that pass through node $v$. In \textbf{Eigenvector Centrality}, the equation $Ax = \lambda x$ is solved, where $A$ is the adjacency matrix and $\lambda$ is the eigenvalue associated with the eigenvector $x$. For \textbf{Katz Centrality} $C_K(v)$, $v$ represents a node, $A^i$ is the i-th power of the adjacency matrix of the graph ($A_{ij}$ represents an edge between the i-th and j-th node), $\alpha$ is the attenuation factor.}
    \begin{tabular}{llp{65mm}}
        \toprule
        \textbf{Metric} & \textbf{Formula} & \textbf{Description}\\
        \midrule
        Degree Centrality~\cite{kenett2015networks} & $C_D(v) = \deg(v)$ & Counts the number of direct connections a node has. Useful for identifying hubs or peripheral nodes.\\
        Betweenness Centrality~\cite{freeman1977set} & $C_B(v) = \sum_{s \neq v \neq t} \frac{\sigma_{st}(v)}{\sigma_{st}}$ & Quantifies how often a node appears on shortest paths between other nodes. Useful for identifying nodes that act as bridges or control points in the network.\\
        Eigenvector Centrality~\cite{bonacich1972factoring} & $C_E(v) = A_v$, where $Ax = \lambda x$ &Assigns relative importance to nodes based on the principle that connections to highly connected nodes contribute more to a node’s score. Indicates influence within densely connected regions. \\
        Katz Centrality~\cite{katz1953new} & $C_K(v) = \sum_{i=1}^{\infty} \alpha^i A^i v$ & Measures node influence by summing over all walks in the graph, with longer walks exponentially damped. Unlike eigenvector centrality, it assigns a non-zero baseline importance to all nodes, allowing even low-degree or peripheral nodes to have influence if they are connected—directly or indirectly—to central nodes. \\
        \bottomrule
    \end{tabular}
    \label{tab:graph_metrics}
\end{table}

\subsubsection{Steiner tree construction}
\label{sec:steiner_tree}

The Steiner tree is computed using Algorithm~\ref{alg:build_steiner}, proposed by Kou et al.~\cite{kou1981fast}. This algorithm approximates the Steiner tree by first computing a complete graph $G_1$, constructed using only the terminal nodes. Then, each edge $(n_i, n_j)$ in $G_1$ is replaced by the corresponding shortest path found in $G$, with all intermediate nodes along these paths also added in. Between each step, the minimality of the graph is ensured by taking the minimum spanning tree of the graph.

While more accurate or theoretically optimal algorithms exist, they are significantly more complex and computationally intensive. Given that the objective of this work is to obtain a structurally meaningful yet computationally tractable representation--rather than an exact minimal tree--Kou et al.'s method offers a suitable trade-off between efficiency and approximation quality.

\begin{algorithm}[tb]
  \DontPrintSemicolon
  \caption{Compute Steiner Tree from a graph and the terminal nodes in that graph, proposed by Kou et al.~\cite{kou1981fast}. \label{alg:build_steiner}}

  \SetKwData{Graph}{graph}
  \SetKwData{Tree}{tree}
  \SetKwData{G}{G}
  \SetKwData{TerminalNodes}{terminalNodes}

  \SetKwInOut{Input}{Input}
  \SetKwInOut{Output}{Output}

  \SetKwData{GOne}{$G_1$}
  \SetKwData{Gs}{$G_s$}
  \SetKwData{TOne}{$T_1$}
  \SetKwData{Ts}{$T_s$}
  \SetKwData{EmptyGraph}{Empty Graph}
  \SetKwData{Node}{Node}

  \Input{\G \texttt{// Graph}, \\
         \TerminalNodes \texttt{// NodeSet}}
  \Output{\Ts \texttt{// TreeGraph}}

  \SetKwFunction{CompleteGraph}{CompleteGraph}
  \SetKwFunction{MinimumSpanningTree}{MinimumSpanningTree}
  \SetKwFunction{GetShortestPath}{GetShortestPath}
  \SetKwFunction{AddNode}{AddNode}
  \SetKwFunction{AddEdge}{AddEdge}
  \SetKwFunction{RemoveNode}{RemoveNode}
  \SetKwFunction{Degree}{Degree}

  \GOne $\leftarrow$ \CompleteGraph{\TerminalNodes}.\;
  \GOne $\leftarrow$ \MinimumSpanningTree{\GOne}.\; \label{lab:mst1}
  \Gs $\leftarrow$ \EmptyGraph.\;
  \ForEach{$(n_i, n_j) \in Edges(G_1)$} {
      $p \leftarrow$ \GetShortestPath($G$, $n_i$, $n_j$) \; \label{alg:shortest_path}
      \ForEach{$(n_k, n_l) \in p$} {
          \AddNode($G_s$, $n_k$) \;
          \AddNode($G_s$, $n_l$) \;
          \AddEdge($G_s$, $n_k$, $n_l$) \;
      }
  }
  \Ts $\leftarrow$ \MinimumSpanningTree{\Gs}.\; \label{lab:mst2}
  \ForEach{$n \in \Ts$} {
      \If{\Degree($n$) = 0} {
      \RemoveNode(\Ts, $n$) \;
      }
  }
\end{algorithm}

\subsubsection{Find rooted trees}
\label{sec:find_roots}

The output of the previous stage is an unrooted tree--a connected, acyclic graph without a designated root. However, the subsequent simplification procedure operates on rooted trees, traversing the structure in a parent-to-child direction. To support this requirement, the unrooted tree must first be converted into one or more rooted trees. Algorithm~\ref{alg:graph_to_tree} formalizes this process, drawing inspiration from Gansner et al.~\cite{gansner1993technique}.

The algorithm iteratively selects a root node from the set of unvisited nodes and constructs a rooted tree by traversing the graph in a depth-first manner from that node. The root is chosen using a selection function that identifies, among the unvisited nodes, the one that maximizes the desired metric. This process is repeated until all nodes in the input graph have been assigned to a tree.

A critical aspect of this procedure is the selection of an initial root node, as defined in step~\ref{alg:select_start_node} of Algorithm~\ref{alg:graph_to_tree}. To produce balanced trees that preserve the features of the underlying geometry, root nodes should be structurally central or functionally significant. This selection can be achieved through a graph metric, such as those discussed in Section~\ref{sec:graph_metrics}. For the purposes of this study, betweenness centrality was used since it provided the most structurally meaningful results for root node extraction in both validation scenarios.

\begin{algorithm}[tb]
  \DontPrintSemicolon
  \caption{Convert the graph to a set of trees. \textit{SelectRootNode} is the function that selects the node from the unvisited nodes that maximizes the desired metric. \label{alg:graph_to_tree}}

  \SetKw{Break}{break}
  \SetKw{Continue}{continue}

  \SetKwData{Graph}{graph}
  \SetKwData{Tree}{tree}
  \SetKwData{Trees}{trees}
  \SetKwData{G}{G}
  \SetKwData{TerminalNodes}{terminalNodes}

  \SetKwInOut{Input}{Input}
  \SetKwInOut{Output}{Output}

  \Input{\Graph \texttt{// Graph}}
  \Output{\Trees \texttt{// TreeSet}}

  \SetKwData{Node}{node}
  \SetKwData{Nodes}{nodes}
  \SetKwData{Visited}{visited}
  \SetKwData{Nil}{nil}
  \SetKwData{RootNode}{rootNode}
  \SetKwData{StartNode}{startNode}
  \SetKwFunction{SelectRootNode}{SelectRootNode}
  \SetKwFunction{BuildTreeFromNode}{BuildTreeFromNode}
  \SetKwFunction{AddTreeNodesFromGraphNodes}{AddTreeNodesFromGraphNodes}
  \SetKwFunction{Neighbours}{Neighbours}
  \SetKwFunction{New}{New}
  \SetKwFunction{Add}{Add}

  \Trees $\leftarrow$ $\emptyset$.\;
  \Visited $\leftarrow$ $\emptyset$.\;

  \While{$\Visited \neq \Graph_{\Nodes}$}{
    \RootNode $\leftarrow$ \SelectRootNode(\Graph, \Visited).\; \label{alg:select_start_node}
    \If {\RootNode $=$ \Nil} {
      \Break \;
    }
    \Tree $\leftarrow$ \BuildTreeFromNode(\Graph, \RootNode, \Visited).\;
    \Trees $=$ \Trees $+$ \Tree.\;
  }
  \Return{\Trees}\;

  \BlankLine
  \SetKwProg{SubAlgo}{Sub-algorithm}{:}{}

  \SubAlgo{\BuildTreeFromNode{\Graph, \StartNode, \Visited}}{
    \If {$\StartNode \in \Visited$} {
      \Return{\Nil}
    }
    \Visited $\leftarrow$ \Visited $+$ \StartNode.\;
    \Tree $\leftarrow$ \New(\StartNode).\;

    \AddTreeNodesFromGraphNodes(\Graph, \Tree, \StartNode, \Visited).\;

    \Return{\Tree}\;
  }

  \SubAlgo{\AddTreeNodesFromGraphNodes{\Graph, \Tree, \Node, \Visited}}{
    \ForEach{$N \in \Neighbours(\Graph, \Node$)} {
      \If {$N \in \Visited$} {
        \Continue
      }
      \Visited $\leftarrow$ \Visited $+ N$.\;
      \Add(\Tree, \Node, $N$).\;
      \AddTreeNodesFromGraphNodes(\Graph, \Tree, $N$, \Visited).\;
    }
  }

\end{algorithm}

\subsubsection{Simplification}
\label{sec:simplification}

The goal of simplification is to reduce complexity by removing redundant nodes without compromising connectivity or spatial coherence. Algorithm~\ref{alg:simplify} achieves this by considering any chain of three connected nodes $A \rightarrow B \rightarrow C$ and evaluating whether the intermediate node $B$ can be safely removed. A predicate $\mathcal{P}(A, B, C)$, described in the following paragraph, determines whether node $B$ is structurally essential or can be omitted without significantly altering geometric or topological characteristics.
Algorithm~\ref{alg:simplify} iterates through the tree, removing any intermediate node $B$ positioned between nodes $A$ and $C$, provided that a direct connection between $A$ and $C$ satisfies $\mathcal{P}(A, B, C)$. If $B$ has multiple child nodes $C_i$, removal of $B$ is permitted only if $\mathcal{P}(A, B, C_i)$ holds for each child connection $A \rightarrow B \rightarrow C_i$. If this condition is not satisfied, the node is retained, preserving structural integrity.
If the previous step (Section~\ref{sec:find_roots}) yields multiple trees, this simplification procedure must be applied individually to each tree.

The predicate $\mathcal{P}(A, B, C)$ used in this study, given by Algorithm~\ref{alg:can_simplify_algo}, verifies the LoS between nodes $A$ and $C$ over the navigation mesh surface. Additionally, it ensures that node $B$ is retained if node $C$ is outside the navigation mesh, a scenario that can occur when the entry/exit method is employed for terminal point identification. This sidesteps the fact that the LoS test on the navigation mesh surface is only meaningful if both points lie on the navigation mesh. Furthermore, the predicate enforces an angular constraint by requiring the angle between the surface normals at nodes $A$ and $B$ to remain within a predefined threshold. This angular criterion prevents simplifications that would distort the structure relative to the original surface geometry.

\begin{algorithm}[tb]
  \DontPrintSemicolon
  \caption{Simplify a tree by removing unnecessary connections.\label{alg:simplify}}

  \SetKw{Break}{break}
  \SetKw{Continue}{continue}

  \SetKwData{Graph}{graph}
  \SetKwData{Tree}{tree}
  \SetKwData{Trees}{trees}
  \SetKwData{G}{G}
  \SetKwData{TerminalNodes}{terminalNodes}
  \SetKwData{RootNode}{rootNode}

  \SetKwInOut{Input}{Input}
  \SetKwInOut{Output}{Output}

  \Input{\Tree \texttt{// Tree}}
  \Output{\Tree \texttt{// Tree}}

  \SetKwData{A}{nodeA}
  \SetKwData{AnyChange}{anyChange}
  \SetKwData{B}{nodeB}
  \SetKwData{C}{nodeC}
  \SetKwData{Children}{children}
  \SetKwData{CanMove}{canMove}
  \SetKwData{Root}{root}
  \SetKwData{StartNode}{startNode}
  \SetKwFunction{SimplifyFromNode}{SimplifyFromNode}
  \SetKwFunction{IsLeaf}{IsLeaf}
  \SetKwFunction{CanSimplify}{CanSimplify}

  \SimplifyFromNode(\Tree, $\Tree_{\Root}$).\;

  \BlankLine
  \SetKwProg{SubAlgo}{Sub-algorithm}{:}{}
  \SubAlgo{\SimplifyFromNode{\Tree, \A}}{
    \If {\IsLeaf(\A)} {
      \Return
    }
    \AnyChange $\leftarrow$ $\text{true}$\;
    \While{\AnyChange}{
        \AnyChange $\leftarrow$ $\text{false}$\;
        \ForEach{$\B \in \A_{\Children}$} {
          \If {\IsLeaf(\B)} {
            \Continue
          }
          \CanMove $\leftarrow$ $\text{true}$\;
          \ForEach{$\C \in \B_{\Children}$} {
            \CanMove $\leftarrow$ \CanMove and $\mathcal{P}(\A, \B, \C)$\; \label{alg:can_simplify}
          }
          \If {\CanMove} {
            \AnyChange $\leftarrow$ $\text{true}$\;
            \ForEach{$\C \in \B_{\Children}$} {
              $\A_{\Children}$ = $\A_{\Children} + \C$\;
            }
            $\A_{\Children}$ = $\A_{\Children} - \B$\;
            \RemoveNode(\Tree, \B).\;
            \Break.\;
          }
        }
    }
  }
\end{algorithm}

\begin{algorithm}[tb]
  \DontPrintSemicolon
  \caption{Simplification acceptance function for the navigation structure extraction use case.\label{alg:can_simplify_algo}}

  \SetKw{Break}{break}
  \SetKw{Continue}{continue}

  \SetKwData{Graph}{graph}
  \SetKwData{Tree}{tree}
  \SetKwData{Trees}{trees}
  \SetKwData{A}{A}
  \SetKwData{B}{B}
  \SetKwData{C}{C}
  \SetKwData{CanCollapse}{canCollapse}
  \SetKwData{TerminalNodes}{terminalNodes}

  \SetKwInOut{Input}{Input}
  \SetKwInOut{Output}{Output}

  \Input{\A, \B, \C \texttt{// Nodes}}
  \Output{\CanCollapse \texttt{// Boolean}}

  \SetKwFunction{HasSurfaceLOS}{HasSurfaceLOS}
  \SetKwFunction{Angle}{Angle}
  \SetKwFunction{IsLeaf}{IsLeaf}
  \SetKwFunction{OnNavMesh}{OnNavMesh}
  \SetKwFunction{Exit}{Exit}

  \tcp{If node C is a leaf, it has to be on the navigation mesh}
  \If {\IsLeaf(\C) and not \OnNavMesh(\C)} {
    \CanCollapse = false
  }

  \tcp{Check if surface normals at points A and B point towards the same direction}
  \ElseIf {\Angle($\A_{normal}$, $\B_{normal}$) > \textit{tolerance}} {
    \CanCollapse = false
  }

  \tcp{There's LoS on the surface of the navigation mesh}
  \ElseIf {\HasSurfaceLOS($\A_{pos}$, $\C_{pos})$} {
    \CanCollapse = false
  }

  \CanCollapse = true;
\end{algorithm}

\subsection{Implementation}
\label{sec:implementation}

The overall process of Algorithm~\ref{alg:gen_structure} is implemented in C\# using the Unity game engine~\cite{unity3d}. Since Unity’s built-in navigation mesh system does not provide direct access to its internal data structures, such as raw polygonal representations, the navigation mesh is instead generated using the DotRecast library~\cite{DotRecast}, a C\# port of the Recast library~\cite{mononen2009navigation}.
Eigenvalues and eigenvectors are computed with the Math.NET Numerics library~\cite{MathNET}.

The Minimum Spanning Tree (MST) in steps \ref{lab:mst1} and \ref{lab:mst2} of Algorithm~\ref{alg:build_steiner} is computed using Kruskal’s algorithm~\cite{kruskal1956shortest} . However, alternative MST algorithms could be employed without loss of generality. In cases where multiple MSTs exist, one is selected arbitrarily.

The full implementation is available under the MIT license at \url{https://github.com/VideojogosLusofona/Graph2Structure}.

\subsection{Test cases and validation}
\label{sec:validation}

Two test cases are used to validate the proposed algorithm. The first consists of extracting the navigation structure of a 3D tile element, as detailed in Subsection~\ref{sec:map_extraction}. The second test case, described in Subsection~\ref{sec:map_analysis}, focuses on player activity analysis. The same process is followed in both cases, with the exception of step 2 of Algorithm~\ref{alg:gen_structure}, i.e., in the way terminal nodes are calculated: in the first case these are obtained through entry/exit points, while in the second case leaf nodes are used for this purpose.

\subsubsection{Map piece structure extraction}
\label{sec:map_extraction}

\begin{figure}[tb]
    \centering
    \Description{The different steps on structure extraction. First we have the piece geometry, then we have the entry/exit points displayed on top of the piece geometry, and finally we have the extracted structure for visualization.}
    \subfloat[]{
        \includegraphics[width=0.3\textwidth]{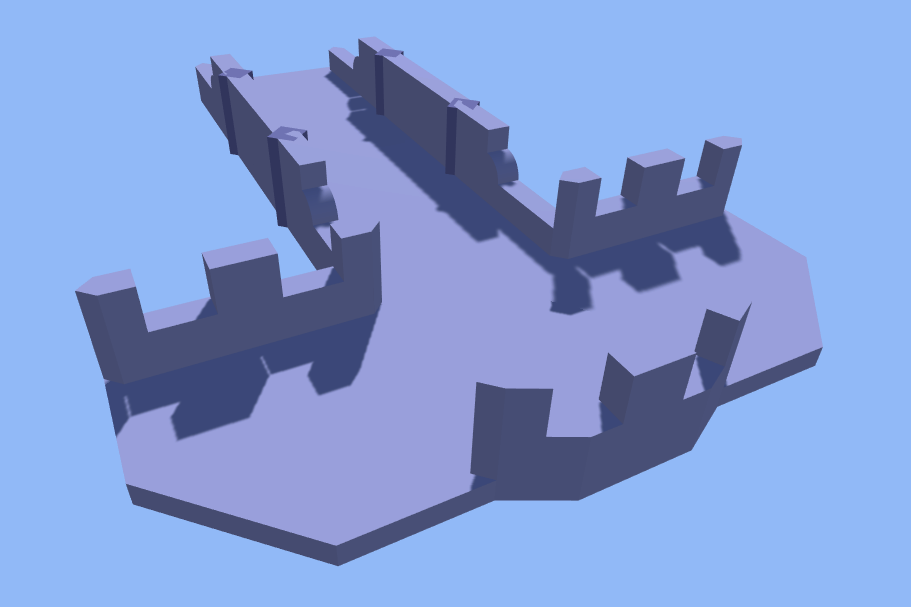}
        \label{fig:extraction_a}}
    \subfloat[]{
        \includegraphics[width=0.3\textwidth]{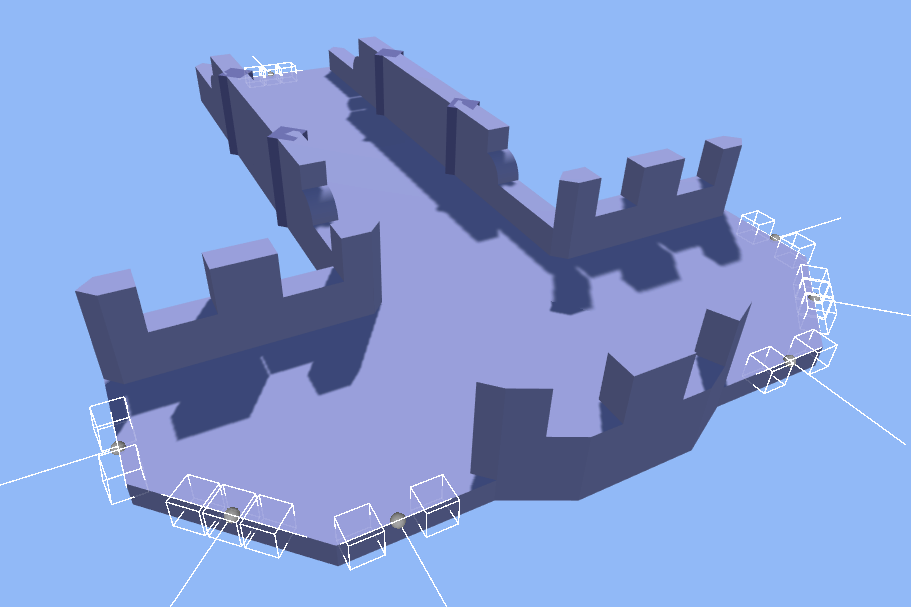}
        \label{fig:extraction_b}}
    \subfloat[]{
        \includegraphics[width=0.3\textwidth]{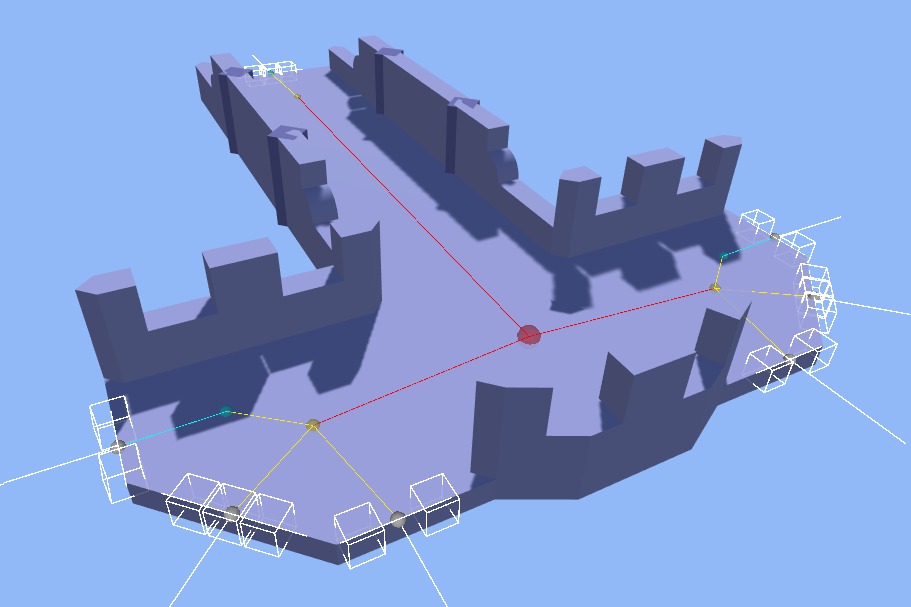}
        \label{fig:extraction_c}}
    \caption{Navigation structure extraction: a) map piece geometry; b) entry/exit points; and, c) extracted structure. The red sphere represents the root of the tree structure, line color represents the depth of the tree (red is depth 1, yellow is depth 2, and so forth).}
    \label{fig:extraction}
\end{figure}

The first test case consists of extracting the navigation structure of a 3D tile element, as illustrated in Fig.~\ref{fig:extraction}. The goal is to generate a simplified structure that preserves the overall shape of the input geometry while encompassing the primary paths between entry/exit points. The resulting structure can then be used to find geometry elements that can fit a specific section of a procedural map, or to facilitate the deformation of the geometry along the navigation routes.

\begin{figure}[tb]
    \centering
    \Description{The effect of simplification (before and after). After, we see a reduction of segments without loss of structural information.}
    \subfloat[]{
        \includegraphics[width=0.45\textwidth]{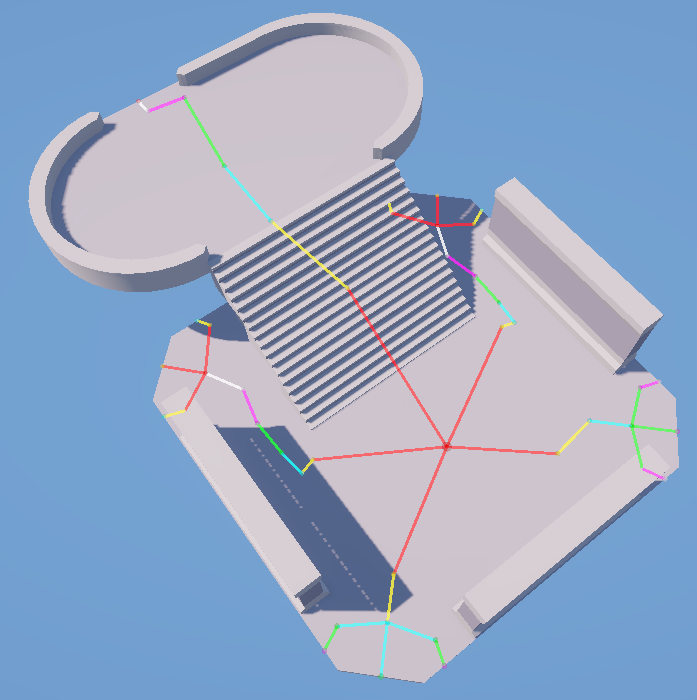}
        \label{fig:simplify01}}
    \subfloat[]{
        \includegraphics[width=0.45\textwidth]{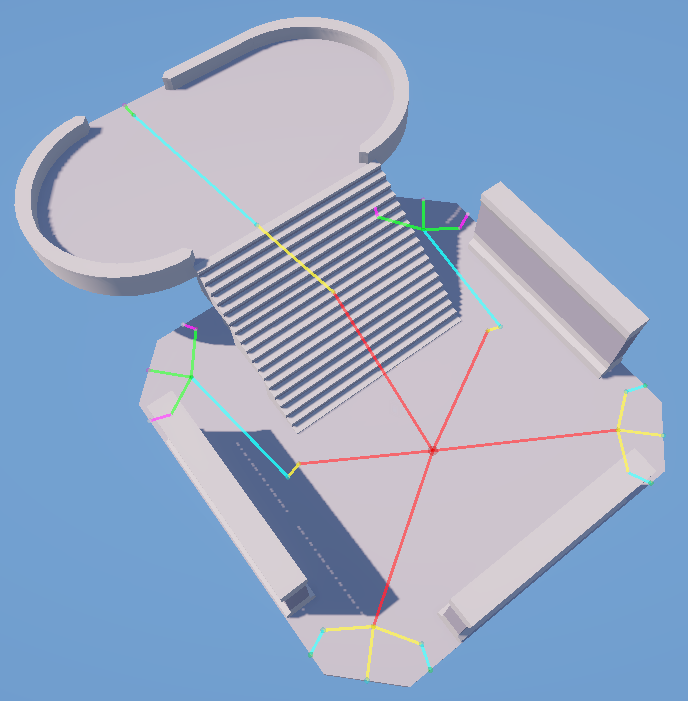}
        \label{fig:simplify02}}
    \caption{Navigation structure extraction: a) no simplification, and, b) after simplification. The red sphere represents the root of the tree structure, line color represents the depth of the tree (red is depth 1, yellow is depth 2, and so forth).}
    \label{fig:simplify_a}
\end{figure}

In this scenario, the initial graph (step~\ref{alg:init_graph} in Algorithm~\ref{alg:gen_structure}) is generated from the navigation mesh of the input geometry. The terminal points (step~\ref{alg:terminal_nodes}) are defined by the entry and exit points using the method described in reference~\cite{de2023automated}.
The process of finding rooted trees (step~\ref{alg:select_start_node} in Algorithm~\ref{alg:graph_to_tree}) is guided by a principle of structural centrality, motivating the use of centrality metrics from graph theory. Among several centrality measures evaluated experimentally, betweenness centrality was selected due to its consistent results across the test set. In cases where multiple nodes exhibit identical betweenness centrality values, a tie-breaking function is applied, selecting the node closest to the geometric center of the object.
The subsequent simplification phase (Figure~\ref{fig:simplify_a}) results in a reduction in both nodes and edges, while preserving the essential topological and navigational structure of the graph.

\subsubsection{Map analysis}
\label{sec:map_analysis}

In this use case, the proposed algorithm generates a simplified graph to assist game designers in analyzing potential player activity. The analysis focuses on betweenness centrality as an indicator of activity density within different sections of the map, specifically in the context of a hypothetical first-person shooter game.

\begin{figure}[tb]
    \centering
    \Description{The effect of simplification (before and after). After, we see a reduction of segments without loss of structural information.}
    \subfloat[]{
        \includegraphics[width=0.8\textwidth]{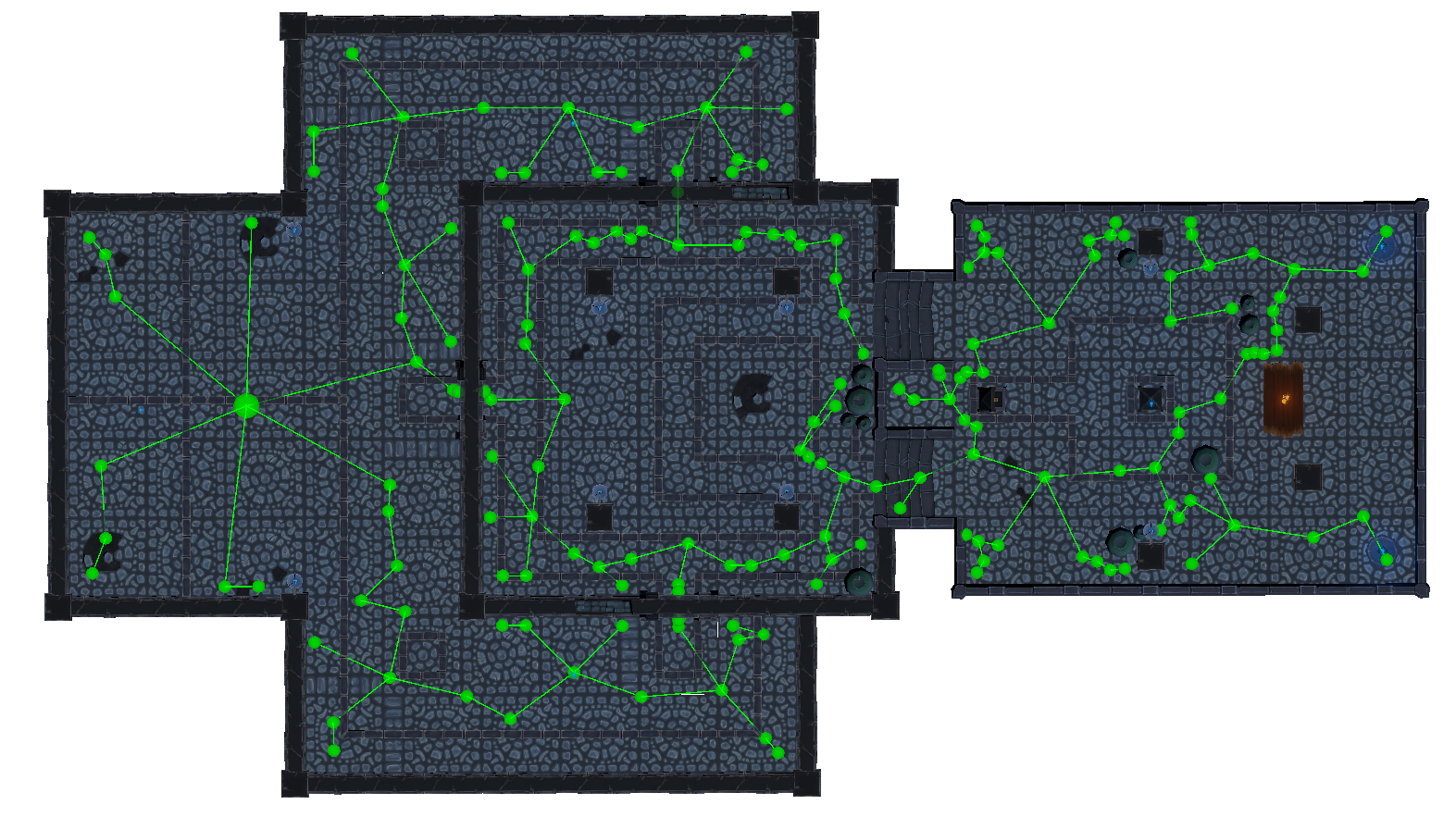}
        \label{fig:simplify03}}
    \\
    \subfloat[]{
        \includegraphics[width=0.8\textwidth]{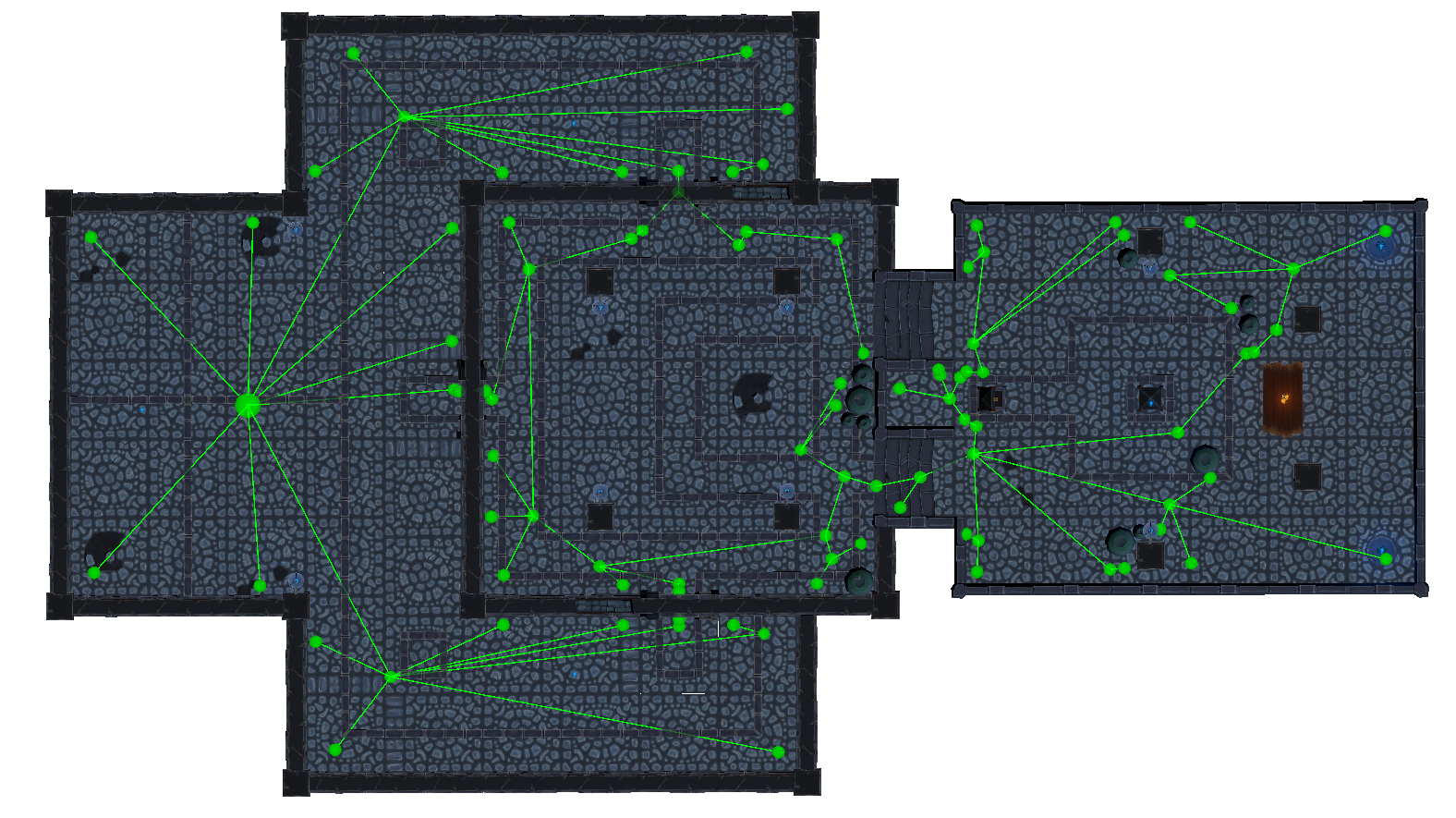}
        \label{fig:simplify04}}
    \caption{Detail of the simplification results on one of the map analysis tests: a) no simplification; and, b) after simplification.}
    \label{fig:simplify_b}
\end{figure}

Similarly to the previous use case, the initial graph (step~\ref{alg:init_graph} in Algorithm~\ref{alg:gen_structure}) is constructed from the navigation mesh of the input geometry. Terminal points (step~\ref{alg:terminal_nodes} in Algorithm~\ref{alg:gen_structure}) are identified using the graph's leaf nodes, as these typically represent dead ends in the map and collectively provide full coverage of the level.
Betweenness centrality is employed at two distinct stages of the process. Like in the previous use case, it is used to select the root nodes for the construction of rooted trees (step~\ref{alg:select_start_node} in Algorithm~\ref{alg:graph_to_tree}), ensuring that the tree structure originates from a strategically central location. After the structure is extracted, it is simplified using Algorithm~\ref{alg:simplify}, reducing the number of nodes and edges in the trees (Fig.~\ref{fig:simplify_b}).

After the process is complete, the analysis stage can begin by converting the trees back into a graph, and computing betweenness centrality to analyze potential player movement patterns. The resulting values are visualized using colored nodes, providing insights into high-traffic areas.

\FloatBarrier

\section{Results}
\label{sec:results}

To evaluate the effectiveness and versatility of the proposed method, the following subsections present the results for the two test cases. In particular, Subsection~\ref{sec:map_extraction_results} demonstrates structural extraction from tile-based geometries, while Subsection~\ref{sec:map_analysis_results} highlights the identification of points of interest and computation of structural metrics within a game level.

\subsection{Map piece structure extraction}
\label{sec:map_extraction_results}

This use case was evaluated using the \textit{Artistic} map pieces from the ``Procedural Generation of 3D Maps With Snappable Meshes'' dataset~\cite{silva2022procedural}. These pieces were originally designed as modular components for procedural map generation and include a variety of corridors and hubs with distinct structural and connectivity properties.

\begin{figure}
    \centering
    \subfloat[]{
        \includegraphics[width=0.34\textwidth]{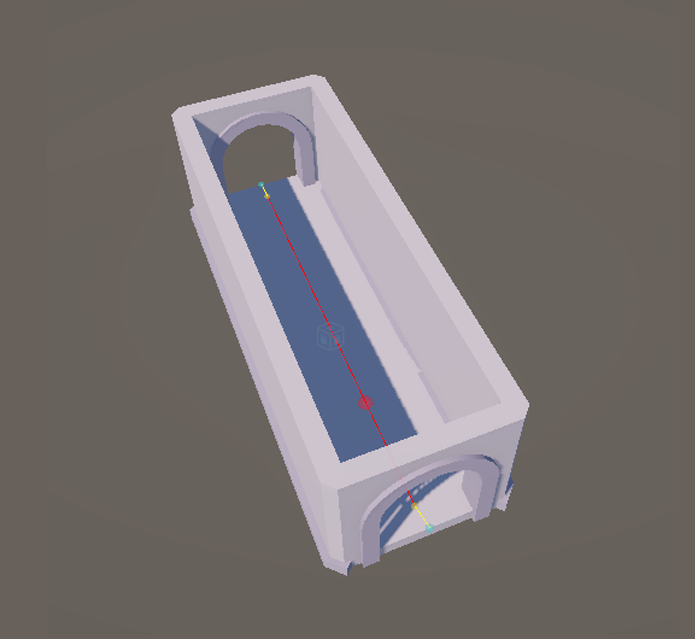}
        \label{fig:corridor1}}
    \subfloat[]{
        \includegraphics[width=0.34\textwidth]{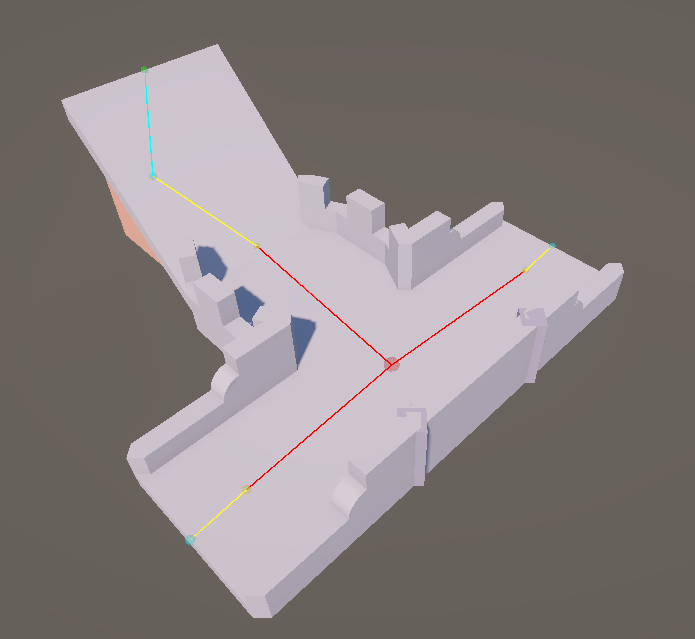}
        \label{fig:corridor5}}
        \\
    \subfloat[]{
        \includegraphics[width=0.34\textwidth]{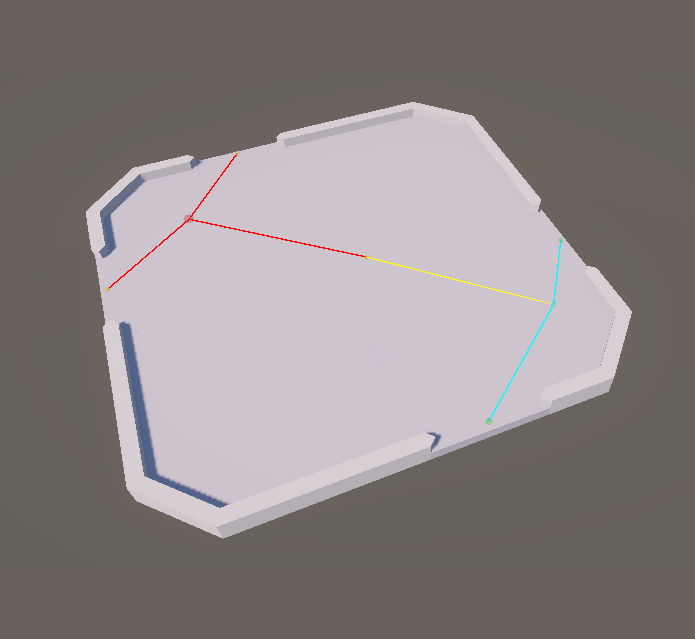}
        \label{fig:hub2}}
    \subfloat[]{
        \includegraphics[width=0.34\textwidth]{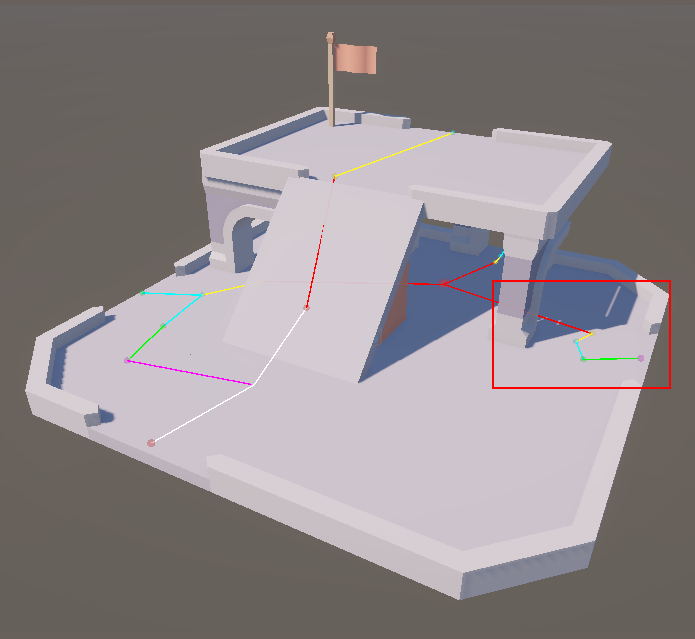}
        \label{fig:hub3}}
        \\
    \subfloat[]{
        \includegraphics[width=0.34\textwidth]{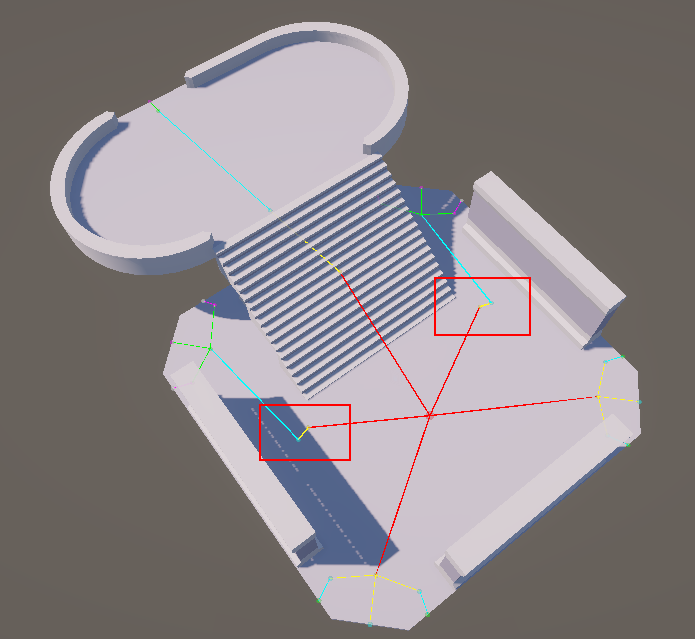}
        \label{fig:hub5}}
    \subfloat[]{
        \includegraphics[width=0.34\textwidth]{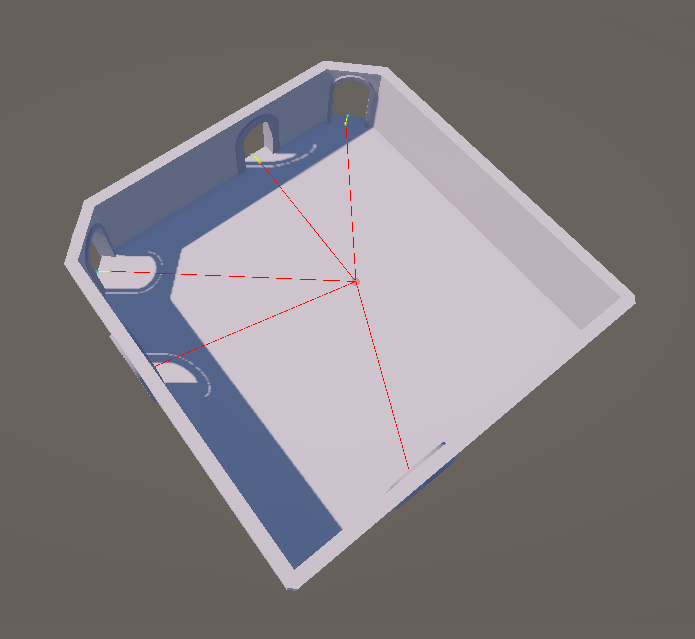}
        \label{fig:hub6}}
    \caption{Map piece structure extraction results: a) corridor 1 map piece; b) corridor 5 map piece, with a ramp; c) hub 2 map piece, correct considering segment minimization, although potentially counter-intuitive, as one might expect a center node connecting to the four exits; d) hub 3 map piece, red rectangle highlight shows a small issue with simplification - green node cannot be removed because the purple node is outside of the navigation mesh; e) hub 5 map piece, red rectangle shows how slopes in the navigation mesh lead to small deviations from expected in the output; and, f) hub 6 map piece. The red sphere represents the root of the tree structure, line color represents the depth of the tree (red is depth 1, yellow is depth 2, and so forth).}
    \Description{Different examples of structure extraction from the map piece, with pieces of different complexity.}
    \label{fig:pieces}
\end{figure}

\begin{figure}[tb]
    \centering
    \includegraphics[width=.5\textwidth]{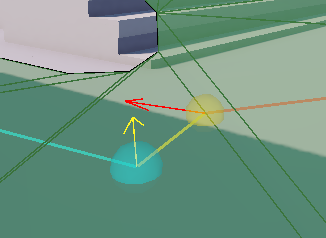}
    \caption{Navigation mesh has an unexpected \textit{bump}, which causes an edge not to be simplified away.}
    \Description{Navigation mesh has an unexpected bump, which causes an edge not to be simplified away.}
    \label{fig:artifact}
\end{figure}

Representative results obtained from applying the proposed approach to different map structures are depicted in Fig.~\ref{fig:pieces}. Results for linear corridors, such as \textit{Corridor 1} (Fig.~\ref{fig:corridor1}) and \textit{Corridor 5}, which includes a ramp (Fig.~\ref{fig:corridor5}), demonstrate that the algorithm can reliably capture the primary navigational paths. However, subtle variations, such as elevation changes, slightly affect the simplification of the extracted paths.
More complex pieces, such as \textit{Hub 2} (Fig.~\ref{fig:hub2}), yield correct results, although they can be counter-intuitive, since one might expect a central node connecting to the four exits. \textit{Hub 3} (Fig.~\ref{fig:hub3}) exhibits minor under-simplification issues. More complex configurations, such as \textit{Hub 5} (Fig.\ref{fig:hub5}), can exhibit limitations in the simplification process. While the overall structure appears correct, closer inspection reveals that a redundant edge was not removed. This behavior is caused by a minor surface irregularity in the navigation mesh, visible in Fig.~\ref{fig:artifact}, where a subtle elevation change leads to a deviation in the normals of adjacent nodes. This normal misalignment prevents the algorithm from identifying the edge as simplifiable. Such artifacts, introduced during navigation mesh generation, can lead to localized inconsistencies in the extracted paths.

In general, the method reliably captures the primary navigational pathways embedded in each map piece, while significantly reducing structural complexity. The resulting trees provide a compact and coherent basis for downstream operations such as geometry-aware deformation and procedural piece matching.

\subsection{Map analysis}
\label{sec:map_analysis_results}

\begin{figure}[tb]
    \centering
    \subfloat[]{
        \includegraphics[width=0.3\textwidth]{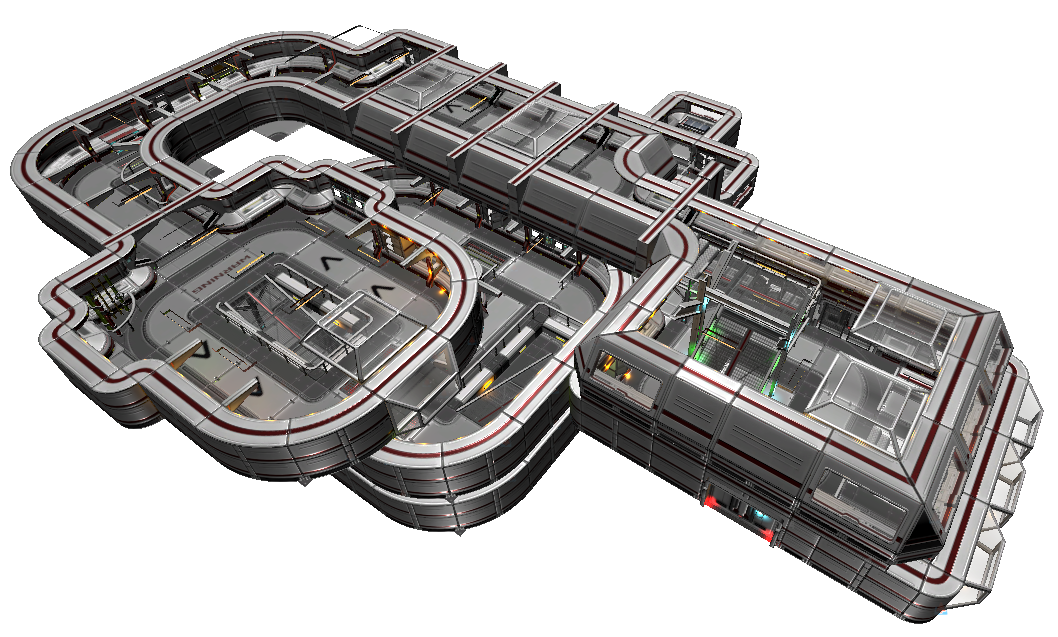}
        \label{fig:2025_01_20_screen01a}}
    \subfloat[]{
        \includegraphics[width=0.3\textwidth]{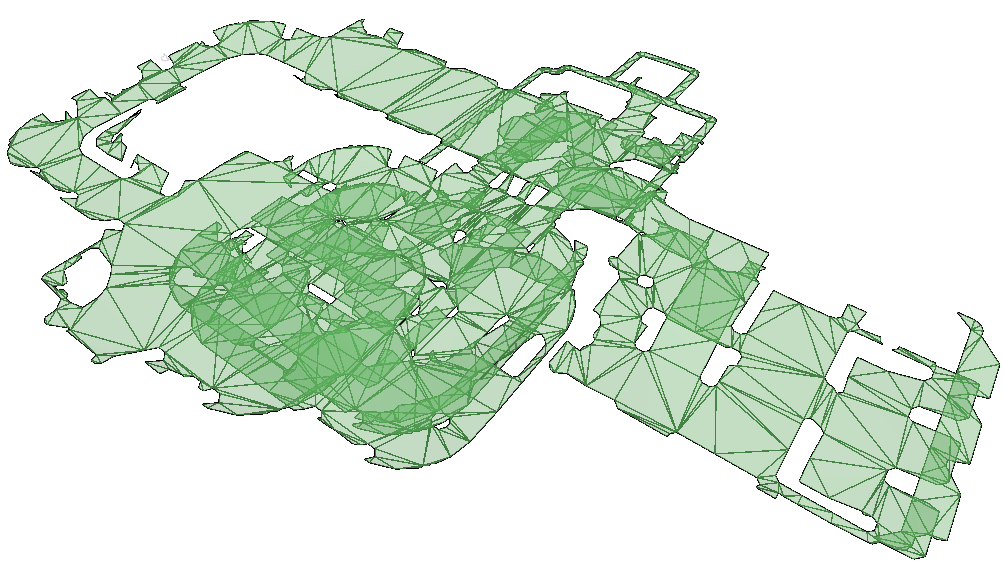}
        \label{fig:2025_01_20_screen01b}}
    \subfloat[]{
        \includegraphics[width=0.3\textwidth]{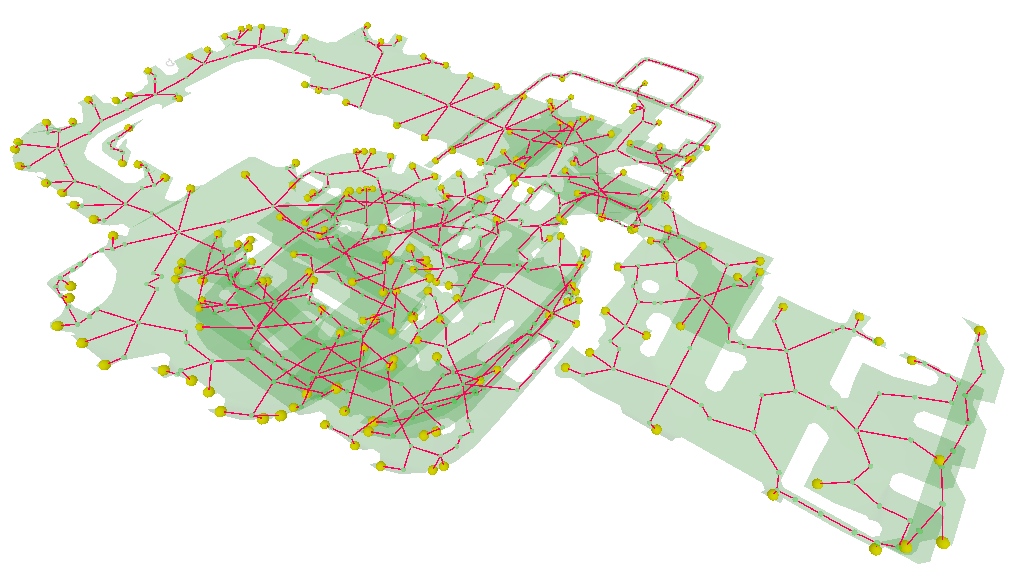}
        \label{fig:2025_01_20_screen01c}}
        \\
    \subfloat[]{
        \includegraphics[width=0.5\textwidth]{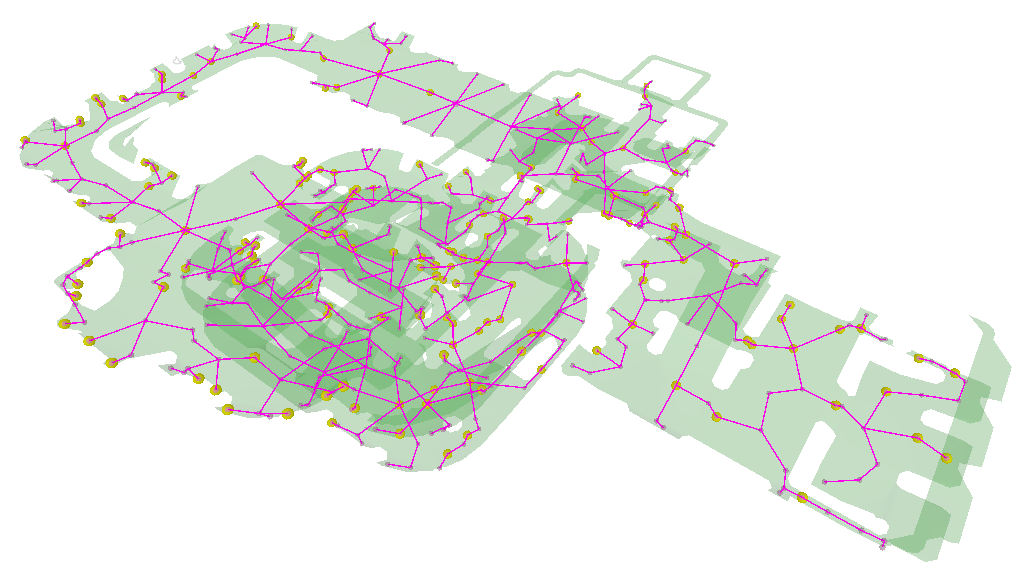}
        \label{fig:2025_01_20_screen01d}}
    \subfloat[]{
        \includegraphics[width=0.5\textwidth]{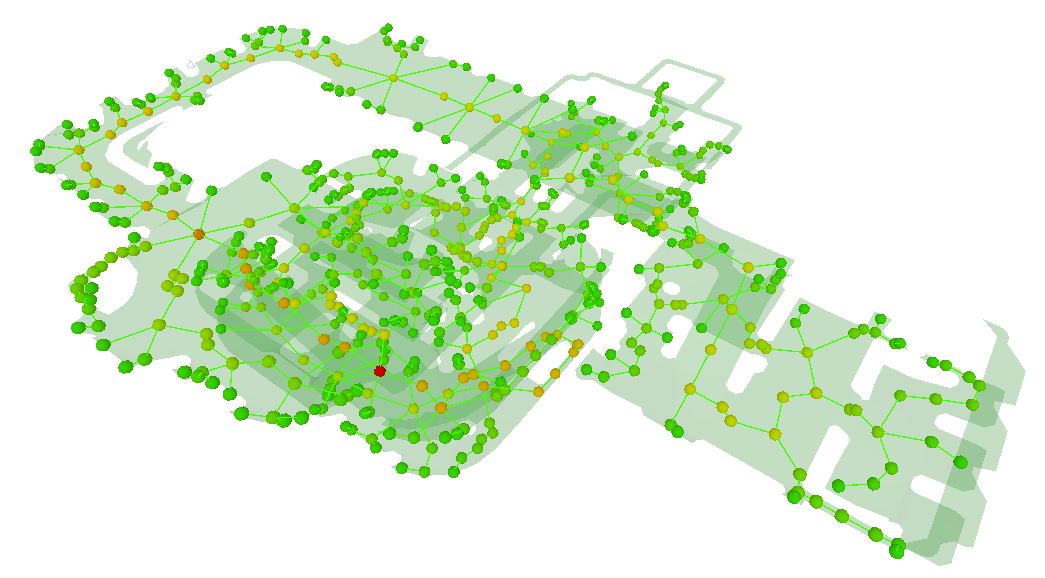}
        \label{fig:2025_01_20_screen01e}}
    \caption{Map analysis using the proposed algorithm: a) test level; b) navigation mesh for the map; c) base graph obtained from the navigation mesh; d) Steiner tree with simplification; and, e) analysis of the betweenness centrality -- green lowest, yellow medium, and red highest.}
    \Description{Different steps on the map analysis algorithm. First we have the test level geometry, then we have the extract navigation mesh. Afterwards, we have the base graph obtained from the navigation mesh, and then the Steiner tree generated from that graph. Finally we have the display of the betweenness centrality on the graph generated by the Steiner tree.}
    \label{fig:2025_01_20_screen01}
\end{figure}

\begin{figure}[tb]
    \centering
    \includegraphics[width=\textwidth]{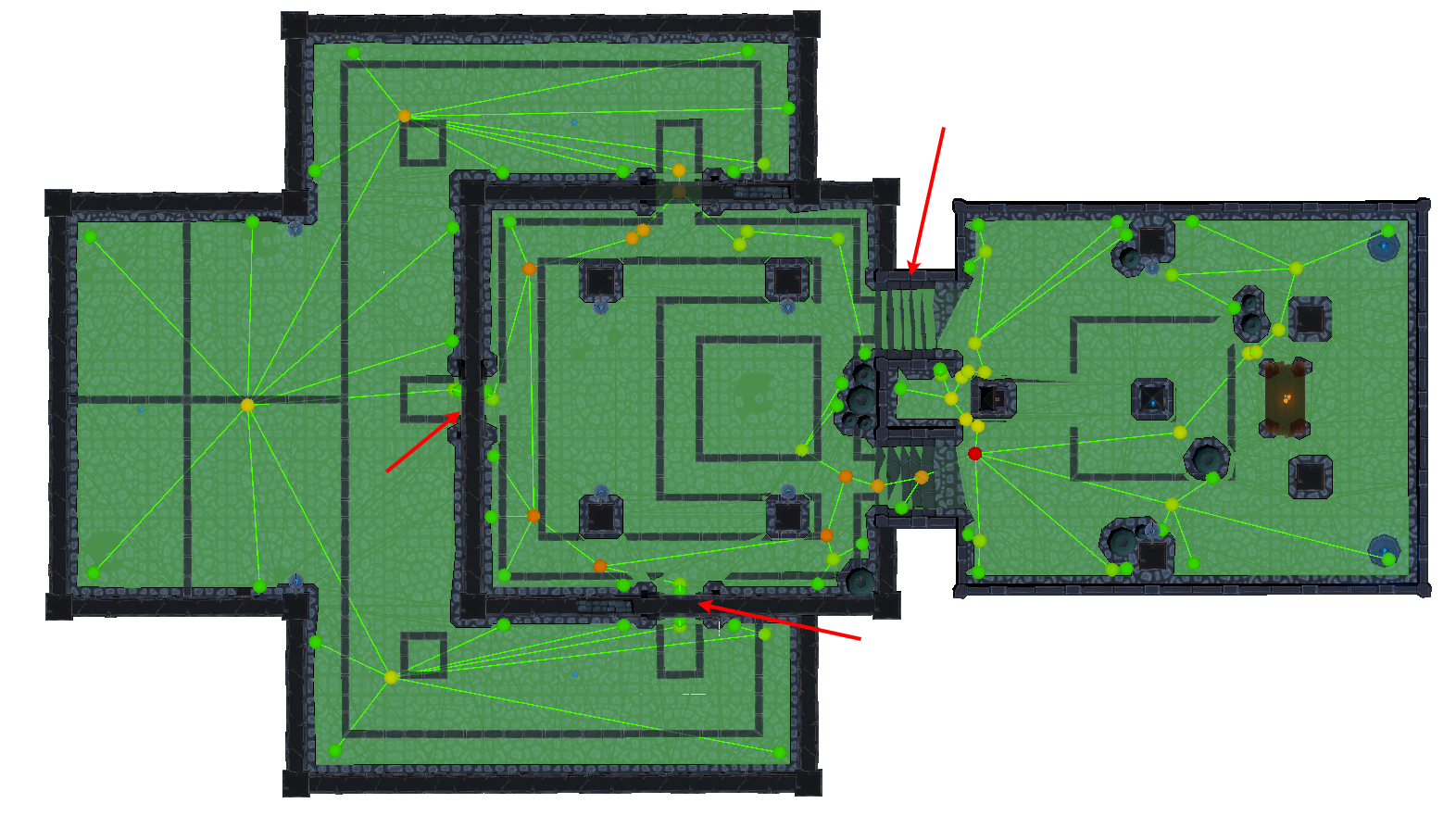}
    \caption{Dungeon map seen from above, showing the hot spots with highest betweenness centrality. The red arrows show areas where the connections are suppressed due to the use of a tree-like structure, which is incapable of handling loops (multiple paths to the same points). The node colors represent the betweenness centrality: green lowest, yellow medium, and red highest.}
    \Description{A dungeon seen from above, overlaid with a graph of the computed structure.}
    \label{fig:dungeon}
\end{figure}

For the second test case, two free asset packs from the Unity Asset Store were used: ``3D Scifi Kit Starter Kit''~\cite{scifikit2024} and ``Blue Dungeon''~\cite{bluedungeon2018}. No modifications were made to these assets.
``3D Scifi Kit Starter Kit'' is a low-poly modular kit that allows users to build sci-fi environments from elements, and it is bundled with a fully assembled spaceship level (Fig.~\ref{fig:2025_01_20_screen01a}), which spans three floors, with multiple corridors leading to different sections and a number of dead ends.
``Blue Dungeon'' is also a low-poly modular pack, focused on fantasy games. It also contains a fully assembled demo level, with a simpler single-floor layout, but composed of highly fragmented geometry with smaller meshes, as shown in Fig.~\ref{fig:dungeon}.

The analysis of the spaceship map reveals that high betweenness centrality values are concentrated in key corridors and intersections, indicating their structural importance within the navigational graph. These results are consistent with intuitive expectations of player movement, which suggest that traversal is more likely to occur through intersection points--particularly those connected to frequently used paths. Among such intersections, those between major areas exhibit disproportionately higher centrality, reflecting their role as strategic choke points.

The analysis of the dungeon map also identifies potential high-traffic areas. However, due to the tree-based nature of the representation, certain features--such as the points highlighted by the red arrows--are excluded from the final result. As Steiner trees do not accommodate loops, such features are not preserved in the simplified structure, limiting the accuracy of the subsequent map analysis in environments where redundant or cyclic paths are relevant.

\subsection{Performance}
\label{sec:performance}

Execution times for each stage of the method are summarized in Table~\ref{tab:timing_structure} and Table~\ref{tab:timing_analysis}. All measurements were obtained in the Unity Editor on Windows 11, using a single thread of an Intel i7-11700K (3.6\,GHz, 8 cores) with 16\,GB RAM.

Each table reports the time required to generate the navigation mesh (\textit{Build NavMesh}) and construct the initial graph (\textit{Build Initial Graph}), along with the number of nodes and edges in this graph. Subsequent columns indicate the time spent identifying terminal nodes (\textit{Identify Terminal Nodes}), the number of terminals selected, Steiner tree computation time (\textit{Compute Steiner Tree}), and tree simplification time (\textit{Simplify Tree}), followed by the size of the final structure and total runtime. Table~\ref{tab:timing_analysis}, which corresponds to full map analysis cases, includes an additional \textit{Post Process} column, capturing the cost of computing betweenness centrality on the final graph. This is the most computationally expensive step in those scenarios.

\FloatBarrier

 % performance_table2
\begin{table}[tb]
    \centering
    \setlength{\tabcolsep}{2pt}
    \caption{Computation time of steps of Algorithm~\ref{alg:gen_structure} for structure extraction, where \textit{t} defines the time in milliseconds for the given step, \textit{\#N} is the number of nodes, \textit{\#E} is the number or edges. $t_{\text{total}}$ represents the total run time in milliseconds. The values of zero are approximate.}
    \label{tab:timing_structure}
    \begin{tabular}{l 
                    rrr  
                    rr   
                    rrr  
                    r    
                    rrr  
                    r}   
        \toprule
        & 
        \multicolumn{3}{c}{\textbf{Step 1}} &
        \multicolumn{2}{c}{\textbf{Step 2}} &
        \multicolumn{3}{c}{\textbf{Step 3}} &
        \multicolumn{1}{c}{\textbf{Step 4}} &
        \multicolumn{3}{c}{\textbf{Step 5}} &
        \\
        \cmidrule(lr){2-4} \cmidrule(lr){5-6} \cmidrule(lr){7-9} \cmidrule(lr){10-10} \cmidrule(lr){11-13} 
        \textbf{Experiment}
        & \textbf{$t$} & \textbf{\#N} & \textbf{\#E} 
        & \textbf{$t$} & \textbf{\#N}
        & \textbf{$t$} & \textbf{\#N} & \textbf{\#E}
        & \textbf{$t$}
        & \textbf{$t$} & \textbf{\#N} & \textbf{\#E}
        & \textbf{$t_{\text{total}}$}\\
        \\
        \midrule
        Corridor 1 Map Piece & \num{363} & \num{12} & \num{11} & \num{0} & \num{2} & \num{0} & \num{8} & \num{7} & \num{0} & \num{0} & \num{5} & \num{4} & \num{363} \\
        Corridor 5 Map Piece & \num{357} & \num{30} & \num{29} & \num{0} & \num{3} & \num{0} & \num{16} & \num{15} & \num{0} & \num{0} & \num{8} & \num{7} & \num{357} \\
        Hub 2 Map Piece & \num{431} & \num{13} & \num{12} & \num{0} & \num{4} & \num{1} & \num{7} & \num{6} & \num{0} & \num{0} & \num{7} & \num{6} & \num{432} \\
        Hub 3 Map Piece & \num{598} & \num{69} & \num{72} & \num{0} & \num{5} & \num{2} & \num{28} & \num{27} & \num{0} & \num{1} & \num{18} & \num{17} & \num{600} \\
        Hub 5 Map Piece & \num{3180} & \num{100} & \num{99} & \num{0} & \num{13} & \num{9} & \num{44} & \num{43} & \num{0} & \num{0} & \num{33} & \num{32} & \num{3189} \\
        Hub 6 Map Piece & \num{406} & \num{40} & \num{39} & \num{0} & \num{5} & \num{1} & \num{25} & \num{24} & \num{0} & \num{0} & \num{11} & \num{10} & \num{407} \\
        \bottomrule
    \end{tabular}
\end{table}

\begin{table}[tb]
    \centering
    \setlength{\tabcolsep}{2pt}
    \caption{Computation time of steps of Algorithm~\ref{alg:gen_structure} for map analysis, where \textit{t} defines the time in milliseconds for the given step, \textit{\#N} is the number of nodes, \textit{\#E} is the number or edges. $t_{\text{total}}$ represents the total run time in milliseconds. The values of zero are approximate.}
    \label{tab:timing_analysis}
    \begin{tabular}{l 
                    rrr  
                    rr   
                    rrr  
                    r    
                    rrr  
                    r    
                    r}   
        \toprule
        & 
        \multicolumn{3}{c}{\textbf{Step 1}} &
        \multicolumn{2}{c}{\textbf{Step 2}} &
        \multicolumn{3}{c}{\textbf{Step 3}} &
        \multicolumn{1}{c}{\textbf{Step 4}} &
        \multicolumn{3}{c}{\textbf{Step 5}} &
        \multicolumn{1}{c}{\textbf{Post Process}} &
        \\ 
        \cmidrule(lr){2-4} \cmidrule(lr){5-6} \cmidrule(lr){7-9} \cmidrule(lr){10-10} \cmidrule(lr){11-13} \cmidrule(lr){14-14} 
        \textbf{Experiment}
        & \textbf{$t$} & \textbf{\#N} & \textbf{\#E} 
        & \textbf{$t$} & \textbf{\#N}
        & \textbf{$t$} & \textbf{\#N} & \textbf{\#E}
        & \textbf{$t$}
        & \textbf{$t$} & \textbf{\#N} & \textbf{\#E}
        & \textbf{$t$} 
        & \textbf{$t_{\text{total}}$}\\
        \midrule
        Spaceship & \num{9719} & \num{769} & \num{784} & \num{7} & \num{214} & \num{37484} & \num{618} & \num{617} & \num{0} & \num{15} & \num{449} & \num{448} & \num{271659} & \num{318891} \\
        Dungeon & \num{70866} & \num{219} & \num{233} & \num{0} & \num{50} & \num{197} & \num{179} & \num{178} & \num{0} & \num{3} & \num{94} & \num{93} & \num{562} & \num{71628} \\
        \bottomrule
    \end{tabular}
\end{table}

In the structure extraction tests (Table~\ref{tab:timing_structure}), computation is dominated by the navigation mesh generation. Once this step is complete, subsequent processing operates on a significantly reduced graph, resulting in very fast runtime for all following stages.

The full map analysis cases (Table~\ref{tab:timing_analysis}) are computationally more demanding. Although the initial graphs for the spaceship and dungeon maps are of similar size, the dungeon scenario requires nearly three minutes for navigation mesh generation, while the spaceship completes this step in under ten seconds. Steiner tree computation time increases with the number of terminal nodes, and post-processing time correlates with the size of the final graph. As a result, despite its shorter initial stages, the spaceship map incurs substantially longer post-processing due to its denser final structure. It is worth noting that in practical applications, the navigation mesh step is often amortized or reused, as it is also required for unrelated gameplay systems such as pathfinding.

\section{Discussion}
\label{sec:discussion}

The results presented in Sections~\ref{sec:map_extraction_results} and~\ref{sec:map_analysis_results} indicate that the proposed method is capable of producing structurally coherent and navigationally meaningful abstractions of complex 3D surfaces. Across a range of geometries--including linear corridors, branching hubs, and complete levels--the extracted trees capture essential connectivity patterns while minimizing redundancy. In the case of tile-based elements, the algorithm retains critical entry/exit relationships and simplifies topological complexity, making it suitable for downstream applications such as modular asset matching, geometry-aware deformation, or procedural layout assembly.

In the larger scale scenarios, the simplified representations can be used as an effective base for structural analysis. Centrality metrics computed over the resulting trees highlight intersections and chokepoints consistent with intuitive notions of traversal density. This supports the method’s use in early-stage level design, spatial diagnostics, or automated evaluation loops. Moreover, the relatively low computational cost--particularly after graph construction--makes the pipeline amenable to integration in real-time or iterative generation workflows.

\begin{figure}[tb]
    \centering
    \includegraphics[width=0.8\textwidth]{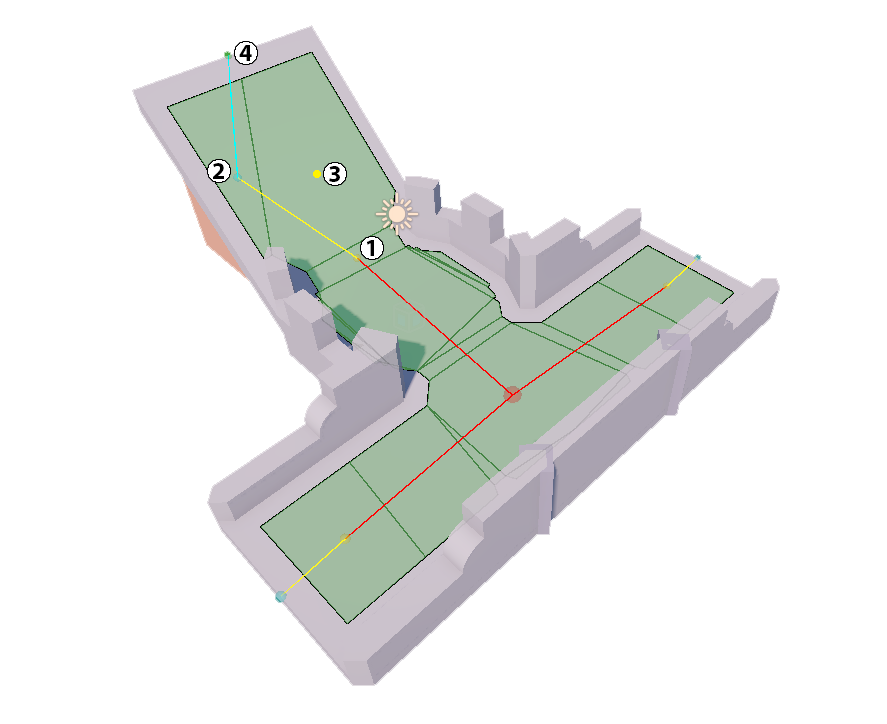}
    \caption{Limitation example due to surface constraints and navigation mesh boundaries. Points $1$, $2$, and $3$ are centers of navigation mesh polygons (green surface), while point $4$ is an entry/exit point located outside the mesh. Although a direct connection from $1$ to $4$ would be structurally preferable, it is prevented by the lack of LoS on the navigation mesh surface and a mismatch in surface normals. As a result, the algorithm requires a detour through points $2$ or $3$, introducing a deviation in the simplified structure.}
    \Description{The structure of a map piece in which one of the arms is twisted}
    \label{fig:limitation1}
\end{figure}

An important factor influencing the effectiveness of the proposed algorithm is the initial graph, as it determines the solution space from which the final tree is derived. Consequently, structural limitations in the initial graph propagate directly to the resulting tree, potentially leading to suboptimal outcomes.
As a first example, in Fig.~\ref{fig:limitation1}, points $1$, $2$ and $3$ are the center point of navigation mesh polygons, and point $4$ was determined using the entry/exit algorithm. The expected direct connection between points $1$ and $4$ is not possible, requiring a detour through either point $2$ or $3$, depending on relative distances. While the simplification algorithm can mitigate such distortions under certain circumstances, it fails in this specific case due to incompatible normal vectors at points $1$ and $4$, and the fact that point $4$ is outside of the navigation mesh. As a result, direct merging of the chain $1 \rightarrow 2 \rightarrow 4$ or $1 \rightarrow 3 \rightarrow 4$ into $1 \rightarrow 3$ becomes impossible. A possible modification involves relaxing surface adherence constraints, permitting edges that deviate slightly from strict conformity with the navigation mesh. In addition, it may be necessary to introduce auxiliary nodes along the boundary of the navigation mesh to bridge connections with points that lie outside its domain. Although these adjustments would introduce minor geometric inaccuracies, such deviations could be considered acceptable for applications where exact surface fidelity is not critical.

Beyond structural limitations, some results exhibit counterintuitive behaviors. For example, Fig.~\ref{fig:hub2} illustrates a situation where an intuitive expectation--a central node connecting directly to multiple exits--is not produced, as this intuitive layout does not correspond to a minimal Steiner tree and results in a greater overall path length. The algorithm instead generates a shorter but less intuitive solution. If a less counter-intuitive structure is preferred, adjustments to the shortest path calculation step (step~\ref{alg:shortest_path} of Algorithm~\ref{alg:build_steiner}) could be explored, for example, by biasing towards collinear segments.

In the context of map analysis, the quality of the results depends significantly on the metric used to evaluate node significance. Betweenness centrality quantifies the proportion of shortest paths that pass through a given node, offering a structural measure of potential traversal importance. However, this metric alone does not necessarily correspond to gameplay-relevant notions such as danger, visibility, or player engagement. In practice, meaningful analysis may require composite metrics that integrate multiple factors—such as coverage, proximity to objectives, or historical heatmap data—rather than relying on structural centrality in isolation. The current system is sufficiently modular to support such extensions, allowing alternative or domain-specific metrics to be incorporated seamlessly into the analysis pipeline.

A more structural limitation stems from the use of tree-based representations, which inherently exclude loops. This is evident in Fig.~\ref{fig:dungeon}, where the regions indicated by red arrows—such as the staircase and side connections—are excluded from the output due to their participation in cyclic paths. Since Steiner trees enforce acyclic connectivity, such features are not retained, reducing the representational fidelity in maps where redundant or multi-path navigation is relevant. This limits the applicability of the method in scenarios requiring complex topologies or resilient routing options.

Nonetheless, the proposed method should be understood as a proof of concept for structure extraction from navigable surfaces, rather than a definitive model of player movement. While the proposed method illustrates how simplified representations can be used to support tasks such as procedural generation and level analysis, it does not constitute a comprehensive model of player behavior. Richer behavioral insights would require integration with empirical data sources—-such as player heatmaps, spatiotemporal clustering, or flow simulations—-which remain outside the scope of the current work.

The algorithm, as presented, is designed to be modular and extensible, supporting multiple avenues for improvement. These include transitioning from strict tree representations to graph-based structures, potentially incorporating Steiner-graph variants with bounded redundancy, or introducing auxiliary constructs to represent loop-like behaviors. Such adaptations can potentially extend the applicability of the method while preserving its computational efficiency and clarity.

\section{Conclusion}
\label{sec:conclusion}

This article presented an approach for extracting simplified structural representations from polygonal surfaces by leveraging Steiner tree approximations and LoS simplification. By transforming surface geometries into graph-based structures, the method achieves a balance between representational fidelity and computational efficiency, making it well suited for PCG, automated spatial analysis, and hierarchical navigation representation.

The proposed method captures essential structural features through a modular pipeline designed for general applicability. While its current formulation is constrained by tree-based representations--which exclude loops and limit support for shortest-path queries—-the architecture is sufficiently flexible to accommodate extensions. However, such adaptations are often use-case dependent and may require rethinking specific stages of the pipeline, such as path selection or simplification, to account for domain-specific requirements or more expressive graph structures.

\begin{acks}
This research was partially funded by the Fundação para a Ciência e a Tecnologia (FCT, \url{https://ror.org/00snfqn58}) under Grants HEI-Lab ref. UIDB/05380/2020, Centro de Tecnologias e Sistemas (CTS) ref. UIDB/00066/2020, and COFAC ref. CEECINST/00002/2021/CP2788/CT0001; and, Instituto Lusófono de Investigação e Desenvolvimento (ILIND, Portugal) under Project COFAC/ILIND/COPELABS/1/2024.
\end{acks}

\bibliographystyle{ACM-Reference-Format}

\end{document}